# Chemical Engineering Journal
## The evolution of nanoparticles due to Brownian coagulation in the temporal mixing layer with AK-iDNS over a long time
--Manuscript Draft--

| | |
|---|---|
| **Manuscript Number:** | |
| **Article Type:** | Research Paper |
| **Section/Category:** | Environmental Chemical Engineering |
| **Keywords:** | population balance equation;  particle size distribution;  moment method;  direct numerical simulation;  average kernel method;  temporal mixing layer |
| **Corresponding Author:** | MINGLIANG XIE, Ph.D.<br>Huazhong University of Science and Technology<br>Wuhan, Hubei CHINA |
| **First Author:** | MINGLIANG XIE, Ph.D. |
| **Order of Authors:** | MINGLIANG XIE, Ph.D. |
| | Kejun Pan |
| **Abstract:** | In this article, the evolution of nanoparticles in a two-dimensional temporal mixing layer over a long time is investigated. the flow field is calculated with direct numerical simulation (DNS), while the particle field is simulated using the average kernel method coupled with iterative direct numerical simulation (AK-iDNS). Under moderate Reynolds number, the flow field undergoes processes of vortex emergence, entrainment, rolling and pairing, merging, and dissipation. Due to the small Stokes number of nanoparticles, and the particles moves closely following the flow field. Meanwhile, the particle undergoes coagulation under the influence of Brownian motion. This article discusses the evolution nanoparticle under the combined effect of advection, diffusion and coagulation. Under the influence of vortices or large-scale coherent structures, the spatial distribution of particle moments is similar to the structure of the flow field. And diffusion and coagulation have a significant impact on the amplitude of the distribution of particle moments. However, diffusion has little impact on the mean distribution, while coagulation has a much greater impact on the mean distribution. As the flow field evolves, the temporal mixing layer degenerates into Couette flow. The particles exhibit similar asymptotic behavior as that of 0-dimensional problem. |

# Cover letter

Dear editor board,

    Now, we submit the second revised article entitled 'The evolution of nanoparticles due to Brownian coagulation in the temporal mixing layer with AK-iDNS over a long time' for consideration by Chemical Engineering Journal.

    In this article, we proposed the AK-iDNS in particle dynamics coupled with DNS in computational fluid dynamics to simulate the evolution of nanoparticle in the fluid flow. Compared with the TEMOM, the present framework has higher efficiency and the same scaling growth rate. The results reveal that particles exhibit similar asymptotic behavior as that of 0-dimensional problem at long time.

    I believe that this manuscript is appropriate for publication by Chemical Engineering Journal.

    I look forward to receiving comments from the reviewers. If you have any queries, please don't hesitate to contact me with the address below.

Yours,
Mingliang Xie
State Key Laboratory of Coal Combustion, Huazhong University of Science and Technology, Wuhan 430074, China
Correspondence Email: mlxie@mail.hust.edu.cn

List of suggested reviewers:

Weber Alfred P. Weber
alfred.weber@mvt.tu-clausthal.de

Jianzhong Lin
mecjzlin@zju.edu.cn

Lian-Ping Wnag
lwang@udel.edu

Zhenjiang You
zhenjiang.you@adelaide.edu.au

Hongwei Wu
h.wu@curtin.edu.cn

Mingzhou Yu
yumingzhou1738@yahoo.com

Thaseem Thajudeen
thaseem@iitgoa.ac.in

**Highlights**

AK-iDNS framework is developed to simulate the evolution of particle in fluid flow;
AK-iDNS has higher efficiency compared to traditional moment method;
The distribution of particle moments is similar to the structures of vortices;
The particles exhibit similar asymptotic behavior as that of 0-dimensional problem at long time;

# The evolution of nanoparticles due to Brownian coagulation in the temporal mixing layer with AK-iDNS over a long time

Mingliang Xie[*]

State Key Laboratory of Coal Combustion, Huazhong University of Science and Technology, Wuhan 430074, China
[*]Corresponding Author's Email: mlxie@mail.hust.edu.cn

Kejun Pan

School of Naval Architecture, Ocean and Energy Power Engineering, Wuhan University of Technology, Wuhan 430063, China

**Abstract:**

In this article, the evolution of nanoparticles in a two-dimensional temporal mixing layer over a long time is investigated. the flow field is calculated with direct numerical simulation (DNS), while the particle field is simulated using the average kernel method coupled with iterative direct numerical simulation (AK-iDNS). Under moderate Reynolds number, the flow field undergoes processes of vortex emergence, entrainment, rolling and pairing, merging, and dissipation. Due to the small Stokes number of nanoparticles, and the particles moves closely following the flow field. Meanwhile, the particle undergoes coagulation under the influence of Brownian motion. This article discusses the evolution nanoparticle under the combined effect of advection, diffusion and coagulation. Under the influence of vortices or large-scale coherent structures, the spatial distribution of particle moments is similar to the structure of the flow field. And diffusion and coagulation have a significant impact on the amplitude of the distribution of particle moments. However, diffusion has little impact on the mean distribution, while coagulation has a much greater impact on the mean distribution. As the flow field evolves, the temporal mixing layer degenerates into Couette flow. The particles exhibit similar asymptotic behavior as that of 0-dimensional problem.

**Keywords:**

population balance equation; particle size distribution; moment method; direct numerical simulation; average kernel method; temporal mixing layer

# Introduction

The population balance equation (PBE) is a mathematical framework utilized to describe the distribution and its dynamics of particles within a disperse system. It is extensively applied in various fields such as chemical engineering, environmental science, and biotechnology to model processes involving particulate systems, including crystallization, polymerization, aerosol dynamics, and biological cell growth (**Friedlander, 2000**). The PBE accounts for the birth, growth, aggregation, and breakage of particles, providing a comprehensive description of how the number and size distribution of particles evolve over time (**Chen et al., 2021**). By incorporating these mechanisms, the equation helps predict the behavior of complex systems where particle interactions play a crucial role. Solving the PBE can yield valuable insights into the design and optimization of industrial processes, ensuring enhanced control over product quality and process efficiency (**Shettigar et al., 2024**). However, it is computationally demanding to directly solve the PBE primarily because of its dependence on the particle volume (**Schumann, 1940**).

Over the past century, various methodologies have been developed in solving the PBE numerically to enhance understanding of particle dynamics evolution. These methodologies encompass the deterministic method of moment, sectional method and the stochastic Monte Carlo method (**Liao and Lucas, 2010**). Given that the particle number density is function of particle volume and time, the numerical methods can be categorized into two primary approaches. The first approach examines the temporal evolution of moments of particle size distribution, represented by the moment method (**Pratsinis, 1988**); the second approach investigate the particle size distribution (PSD) at extended time scales, represented by self-preserving theory (**Friedlander and Wang, 1966**). The sectional method and Monte Carlo method enable simultaneously examination of the evolution of particle size distribution with respect to time and particle volume, and are generally considered direct numerical simulation methods for investigating particle dynamics.

The sectional method divides the particle volume space into several discrete sections or bins. Within each section or bin, the size of particles can be considered as a constant or linearly distributed function, and the PBE can be solved. To achieve the required accuracy, the number of sections needs to reach hundreds or higher, which significantly increases the computational cost. Furthermore, the sectional method requires a preset range of particle size, i.e., the minimum and maximum particle volume; consequently, the volume of new particles formed after coagulation will inevitably exceed the preset maximum. Therefore, the sectional method will inevitably result in numerical divergence as the particles evolves over time. However, due to its high accuracy with a large number of sections, sectional method is often used as a reference for evaluating other methods (**Wang et al., 2007**).

The Monte Carlo method is fundamentally based on the physical model of the research object, and delineates the evolution of particles through stochastic processes and simulated events. Theoretically, it can account for the effect of finite scale, spatial correlation and fluctuations of particles. When conducting Monte Carlo simulations, it is essential to predetermine the total number of particles and their initial distribution. As particles interact, they tend to concentrate towards the central regions of the particle size distribution, while significant deviations occur at the upper and lower extremities. To enhance the accuracy of numerical results, the Monte Carlo method necessitates a substantial number of initial particles and considerable computation resources, which constitute the primarily factors limiting its widespread application in theoretical analysis and engineering practice (**Zhao and Zheng, 2011**).

The moment method transforms the particle size distribution into various moments through integral transformation, thereby converting the PBE into a set of ordinary integra-differential equations. Since the particle size distribution is equivalent to its infinite order moments, and in practical application only a limited number of moment equation can be calculated, the primary challenge for the moment method is the closure problem of the system of moment equations. To address the closure problem, researchers have developed method of moment (MOM) based on the lognormal distribution (**Lee et al., 1984**), quadrature method of moment (QMOM) (**McGraw, 1997**), Taylor series expansion method of moment (TEMOM) (**Yu et al., 2008**), and other approaches. Another significant challenge faced by the moment method is its inverse problem, i.e., how to reconstruct the particle size distribution from a set of finite order moments. Research on the inverse problem of the moment method remains an active area of investigation and a challenge in mathematics and engineering fields (**John et al., 2007**).

Despite these difficulties and challenges, the moment method is increasingly been applied to numerical simulation in particle science and technology due to its characteristics of simplicity, high efficiency and accuracy. It should be noted out that the closure of the moment equations is approached using the Taylor-series expansion technique in the TEMOM. Through constructing a system of the three first-order ordinary differential equations, the most important moments for describing the particle dynamics, including the particle number density, particle mass and geometric standard deviation, are obtained. This approach has no prior requirement for the particle size spectrum, and the limitation inherent in the lognormal distribution theory automatically eliminated. However, when dealing with certain specialized collision kernels (such as the sedimentation kernel), TEMOM cannot accurately obtain the corresponding moment models, and alternative method are required in such cases (**Xie, 2024**).

If the collision frequency function of particles is homogeneous with respect to its variables, then the PBE can be transformed into an ordinary differential equation through similarity transformation. Previous studies have revealed that the particle size distribution will reach a self-preserving form at long times (**Friedlander and Wang, 1966**). Similarity theory provides an analytical method for analyzing the asymptotic behavior of the coagulating system. Recently, Xie proposed an iterative direct numerical simulation (iDNS) to solve the governing equation of self-preserving size distribution, using the asymptotic solution of TEMOM as the initial conditions. This establishes a one-to-one mapping relationship between TEMOM and similarity theory based on iDNS algorithm, laying a foundation for analyzing the mathematical properties and physical characteristics of the PBE (**Xie, 2023**). Furthermore, Pan et al. proposed a concise and fast framework based on the average kernel and iterative direct numerical simulation (AK-iDNS) to solve the PBE with a physically realistic kernel (**Pan et al., 2024**). Compared with the TEMOM, this new framework has the same asymptotic growth rate but in a more concise form, decoupling particle number density from other moments. Using the analytical solution of PBE with average kernel as initial condition, an iteratively corrected similarity solution satisfying the original kernel characteristics is obtained. The results not only meet the theoretical accuracy and efficiency requirements, but also agree with the experimental data. And AK- iDNS is expected to become an upgraded alternative method to TEMOM.

Research on the evolution of nanoparticles in fluid flow is essential for advancing our understanding of multiphase systems and has significant implications across various scientific and engineering disciplines. In fluid dynamics, a temporal mixing layer is a fundamental flow configuration where two parallel streams of fluid with different velocities interact over time (**Chorin A.J., 1968; Chorin A.J., 2000**). The temporal mixing layer is characterized by strong shear and coherent structures. Coherent structures refer to organized patterns of fluid motion that persist over time and space, significantly influencing the overall behavior of the flow. These structures typically manifest as vortices or eddies, which play a pivotal role in the entrainment and mixing of fluid elements. Analyzing the formation, evolution and interaction of these coherent structures provides insights into the mechanisms driving turbulence, energy transfer, and scalar mixing in various engineering and natural systems. Understanding these dynamics is essential for improving predictive models and optimizing processes in fields such as aerodynamics, environmental engineering, and industrial mixing.

Over the past few decades, significant advancements in computational methods and resources have enabled researchers to simulate temporal mixing layers with increasing accuracy and resolution (**Lele, 1992; Strang, 2007**). Early simulations focused on capturing the basic features of mixing layers, such as the growth of the shear layer and the formation of large-scale vortices. With the development of more sophisticated algorithms and an increase in computational power, it has become possible to resolve finer scales of turbulence and capture the intricate details of coherent structures within the flow. Modern numerical techniques, such as direct numerical simulation (DNS), have been particularly successful in providing high-fidelity data on the evolution of temporal mixing layers. However, performing DNS over long time periods presents several significant challenges, such as computational cost, data management, numerical stability and accuracy, physical modeling, parallelization and validation. Addressing these challenges requires advanced computational techniques, efficient algorithms, and substantial computational resources.

The integration of advanced models for particle dynamics, such as PBE, with fluid flow simulations has opened new avenues for studying the interaction between flow and particulate matter. Do the particles exhibit similar asymptotic behavior in the flow field under long-term evolution? This study aims to combine the DNS in computational fluid dynamics with AK-iDNS in particle dynamics to simulate the evolution of nanoparticles in the temporal mixing layer due to Brownian coagulation over a long time. Understanding how nanoparticles distribution and homogenize within a coherent structure can enhance the design of processes that require uniform particle distribution.

## Governing equation

**Fluid fields**

The flow is considered to be the constant density two-dimensional temporal mixing layer containing the nano-scale particles as shown **Figure 1**. The temporal mixing layer can be thought of an approximation to spatially developing mixing layer, and the simulation of the time-developing mixing layer allows a much higher resolution for computation (**Rogers and Moser, 1992; Moser and Rogers, 1993**). For the temporal mixing layer, the periodic boundary conditions can be imposed in the stream-wise direction ($x$ coordinate). Considering that all the perturbations vanish rapidly as $y \to \infty$, where $y$ is the coordinate in the transverse direction, periodic conditions can be also imposed in the transverse direction by introducing the image flows far enough from the mixing layer center. Thus, the standard Fourier pseudo-spectral method can be applied directly (**Schmid and Henningson, 2001; Drain, 2002**). The primary transport variables for the flow field are the fluid velocity and pressure. These variables are governed by the Navier-Stokes equations:

$$\frac{\partial u}{\partial x} + \frac{\partial v}{\partial y} = 0$$
$$\frac{\partial u}{\partial t} + u\frac{\partial u}{\partial x} + v\frac{\partial u}{\partial y} = -\frac{1}{\rho}\frac{\partial p}{\partial x} + \nu\left(\frac{\partial^2 u}{\partial x^2} + \frac{\partial^2 u}{\partial y^2}\right) \quad (1)$$
$$\frac{\partial v}{\partial t} + u\frac{\partial v}{\partial x} + v\frac{\partial v}{\partial y} = -\frac{1}{\rho}\frac{\partial p}{\partial y} + \nu\left(\frac{\partial^2 v}{\partial x^2} + \frac{\partial^2 v}{\partial y^2}\right)$$

Where $\rho$ is the fluid density; $p$ is the pressure; $\nu$ is the kinematic viscosity; $u$, $v$ are the velocity component in the $x$ and $y$ directions, respectively. the initial velocity for the time developing mixing layer consists of the following two parts: the base flow $(U, V)$ and the corresponding disturbances $(u', v')$. The transverse velocity $V$ of base flow is set to zero ($V = 0$) and the streamwise velocity $U$ is specified using a hyperbolic tangent profile:

$$U = \frac{U_1 + U_2}{2} + \frac{U_1 - U_2}{2}\tanh\frac{y}{2\theta} \quad (2)$$

Where $U_1$ and $U_2 (= -U_1)$ are the far field velocities of two parallel streams on each side of the shear layer, and $\theta$ is the initial momentum thickness, and the upper and lower boundaries are slip walls. Using slip walls allows us to concentrate on the shear region without having to resolve the boundary layer on the lower and upper walls.

**Particle fields**

The transport of the nano-scale particles dispersed through the fluid is governed by the particle PBE. The PBE describes the particle dynamics under the effect of different physical and chemical processes: advection, diffusion, coagulation, condensation, nucleation and other internal/external forces (**Friedlander, 2000**). In the present study, only the Brownian coagulation in the free molecule regime is considered, and the PBE can be written as:

$$\frac{\partial n}{\partial t} + u\frac{\partial n}{\partial x} + v\frac{\partial n}{\partial y} = \frac{\partial}{\partial x}\left(D_n\frac{\partial n}{\partial x}\right) + \frac{\partial}{\partial y}\left(D_n\frac{\partial n}{\partial y}\right) + \left[\frac{\partial n}{\partial t}\right]_{coag} \quad (3)$$

where the source term $[\partial n/\partial t]_{coag}$ is given by the classical Smoluchowski equation as

$$\left[\frac{\partial n}{\partial t}\right]_{coag} = \frac{1}{2}\int_0^v \beta(v_1, v - v_1)n(v_1)n(v - v_1)dv_1 - \int_0^\infty \beta(v_1, v)n(v_1)n(v)dv_1 \quad (4)$$

which represents the effects of particle-particle interactions resulting in coagulation, and $n(v,t)dv$ is the number density of particles per unit spatial volume with particle volume from $v$ to $v + dv$ at time $t$; and $\beta$ is the collision kernel of coagulation. For brownian coagulation in the free molecule regime, the collision kernel takes:

$$\beta = \left(\frac{3}{4\pi}\right)^{\frac{1}{6}} \left(\frac{6k_BT}{\rho_p}\right)^{\frac{1}{2}} \left(\frac{1}{v_i} + \frac{1}{v_j}\right)^{\frac{1}{2}} \left(v_i^{\frac{1}{3}} + v_j^{\frac{1}{3}}\right)^2 \tag{4}$$

where $k_B$ is the Boltzmann's constant; $T$ is the fluid temperature; $\rho_p$ is the particle density; and $D_n$ is the particle diffusion coefficient, which is given by Einstein-Smoluchowski relation as

$$D_n = \frac{k_BT}{f} \tag{5}$$

where $f$ is the friction coefficient of particles in fluids. In the free molecule regime, the friction coefficient can be derived from the kinetic theory as (**Epstein, 1924**)

$$f = \frac{2}{3}d_p^2\rho \left(\frac{2\pi k_BT}{m}\right)^{\frac{1}{2}} \left(1 + \frac{\pi\alpha_p}{8}\right) \tag{6}$$

where $m$ is the molecular mass of fluid molecules, $\rho$ is the density of fluid, $\alpha_p$ is the accommodation coefficient, and $d_p$ is the diameter of particles, and $d_p = \left(\frac{6v}{\pi}\right)^{\frac{1}{3}}$.

## Mathematical modeling

### AK-iDNS framework for Smoluchowski equation

#### The average kernel method

To solve the Smoluchowski equation with collision kernel depending on the particle size, Schumann proposed the average kernel method (**Schumann, 1940**), which is defined with Laplace transformation as

$$\int_0^\infty \int_0^\infty \bar{\beta} \exp\left(-\frac{v_1+v}{v_a}\right) dv_1 dv = \int_0^\infty \int_0^v \beta(v_1,v) \exp\left(-\frac{v_1+v}{v_a}\right) dv_1 dv \tag{7}$$

where $v_a$ is the algebraic mean volume of particle size distribution. Through operation and reorganization, the average kernel can be represented as

$$\bar{\beta} = \frac{1}{v_a^2} \int_0^\infty \int_0^v \beta(v_1,v) \exp\left(-\frac{v_1+v}{v_a}\right) dv_1 dv \tag{8}$$

For homogeneous collision kernel, it has the following properties:

$$\begin{cases} \beta(\alpha v, \alpha v_1) = \alpha^q \beta(v,v_1); \\ \beta(v,v_1) = \beta(v_1,v); \\ \beta(v,v_1) \geq 0; \end{cases} \tag{9}$$

in which $q$ is power index, and $\alpha$ is the scale factor. It is easy to prove that the homogeneous collision kernels satisfy the following differential equation:

$$v\frac{\partial\beta}{\partial v} + v_1\frac{\partial\beta}{\partial v_1} - q\beta = 0 \tag{10}$$

If the scale factor is $\alpha = 1/v_a$, the collision kernel can be expressed as

$$\beta(v,v_1) = v_a^q \, \beta\left(\frac{v}{v_a}, \frac{v_1}{v_a}\right) \tag{11}$$

Let the dimensionless particle volume as

$$\eta = \frac{v}{v_a} \tag{12}$$

And the average kernel can be expressed as

$$\bar{\beta} = pv_a^q \tag{13}$$

where $p$ is a proportional factor, and it can be calculated as

$$p = \int_0^\infty \int_0^{v_a\eta} e^{-\eta-\eta_1} \beta(\eta,\eta_1) d\eta_1 d\eta \tag{14}$$

which usually represents the total collision frequency of particle coagulating system. Due to the symmetry of the homogeneous collision kernel, it can be simplified as

$$p = \frac{1}{2}\int_0^\infty \int_0^\infty e^{-\eta-\eta_1} \beta(\eta,\eta_1) d\eta_1 d\eta \tag{15}$$

For Brownian coagulation in the free molecule regime, the parameters $p$ and $q$ can be calculated as (**Pan et al., 2024**)

$$p = 4\sqrt{2}B_1\Gamma\left(\frac{13}{12}\right), \quad q = 1/6 \tag{16}$$

where $\Gamma$ is the Euler gamma function, and the average kernel can be expressed as $\bar{\beta} = 4\sqrt{2}B_1\Gamma\left(\frac{13}{12}\right)v_a^{\frac{1}{6}}$.

**The moment model**

In the moment method, the $k$th order moment of the volume-based PSD is defined as

$$M_k(t) = \int_0^\infty v^k n(v,t)dv \tag{17}$$

Using the moment transformation above, the Smoluchowski equation is transformed into a series of ordinary differential equations as

$$\frac{dM_k}{dt} = -\frac{1}{2}\int_0^\infty \int_0^\infty [(v+v_1)^k - v^k - v_1^k]\beta(v,v_1)n(v)n(v_1)dvdv_1 \tag{18}$$

Generally, the minimum set of moments required to close the moment equations is the first three, i.e., $M_0$, $M_1$ and $M_2$. The zeroth order moment ($M_0$) represents the total particle number concentration; the first moment ($M_1$) is proportional to the particle mass concentration; and the second moment ($M_2$) describes the dispersity of PSD.

Substituting the average kernel $\bar{\beta} = pv_a^q$ into the moment equations, it can be found

$$\begin{cases} \left[\frac{dM_0}{dt}\right]_{coag} = -\frac{1}{2}\bar{\beta}M_0^2 \\ \left[\frac{dM_1}{dt}\right]_{coag} = 0. \\ \left[\frac{dM_2}{dt}\right]_{coag} = \bar{\beta}M_1^2 \end{cases} \tag{19}$$

Usually, $v_a$ is selected as the algebraic mean volume of PSD, its expression can be written as

$$v_a = \frac{M_1}{M_0} \tag{20}$$

For the classical Smoluchowski equation, $M_1$ remains constant due to the rigorous mass conservation requirement, and the moment equations is decoupled. And the evolution of particle number density can be obtained with only one equation. According to the value of index $q$, the solution of zeroth order moment can be divided into three types. In the present study, the index $q = 1/6$, then analytical solution can be found as

$$M_0 = \left[M_{00}^{q-1} - \frac{q-1}{2}pM_1^q(t-t_0)\right]^{\frac{1}{q-1}} \tag{21}$$

The corresponding asymptotic scaling growth rate of the $M_0$ can be found as

$$\frac{1}{M_0}\frac{dM_0}{dt} = \frac{1}{(q-1)t} \tag{22}$$

**Similarity transformation and corrected similarity solution**

The similarity transformation is based on the assumption that the fraction of the particles in a given size range is only volume dependent on the dimensionless particle volume ($\eta = v/v_a$), then the similarity transformation takes the form as (**Friedlander and Wang, 1966**)

$$\begin{cases} \eta = \frac{v}{v_a} \\ \psi(\eta) = \frac{M_1}{M_0^2}n(v,t) \end{cases} \tag{23}$$

in which $\eta$ is dimensionless particle volume, $\psi$ is the dimensionless particle size distribution, and the Smoluchowski equation is transformed into the following governing equation of self-preserving theory (**Xie, 2023**):

$$A\left[2\psi(\eta) + \eta\frac{d\psi(\eta)}{d\eta}\right] = C(\eta) - \psi(\eta)g(\eta) \tag{24}$$

in which the total collision frequency $A$, the gain term $C(\eta)$, and the loss term $g(\eta)$ are given as follows:

$$\begin{cases} A = -\frac{1}{2}\int_0^\infty \int_0^\infty \beta(u\eta, u\eta_1)\psi(\eta)\psi(\eta_1)d\eta d\eta_1 \\ C(\eta) = \frac{1}{2}\int_0^{u\eta} \beta(u\eta_1, u(\eta-\eta_1))\psi(\eta_1)\psi(\eta-\eta_1)d\eta_1 \\ g(\eta) = \int_0^\infty \beta(u\eta_1, u\eta)\psi(\eta_1)d\eta_1 \end{cases}$$

$$\tag{25}$$

The dimensionless distribution function $\psi(\eta)$ to be solved is restricted by the following mathematical and physical constraints:

$$\int_0^\infty \psi(\eta)d\eta = 1 \,;\, \int_0^\infty \eta\psi(\eta)d\eta = 1 \,;\, \psi(\eta) \geq 0 \tag{26}$$

and the boundary conditions:

$$\begin{cases} \psi(\eta) \to 0 \;\; for \;\; \eta \to \infty \\ \eta\psi(\eta) \to 0 \;\; for \;\; \eta \to 0 \end{cases} \tag{27}$$

The existence of the similarity solution is related to the asymptotic properties of the kernel. Substituting the average kernel $\bar{\beta} = pu^q$ into the governing equation, it can be found that

$$\begin{cases} A = -\frac{1}{2}pu^q \\ C(\eta) = \frac{1}{2}pu^q \int_0^{u\eta} \psi(\eta_1)\psi(\eta-\eta_1)d\eta_1 \\ g(\eta) = pu^q \end{cases}$$

$$\tag{28}$$

Therefore, the governing equation can be simplified as

$$\eta \frac{d\psi(\eta)}{d\eta} + \int_0^\eta \psi(\eta_1)\psi(\eta - \eta_1)d\eta_1 = 0 \tag{29}$$

Using the Laplace transformation:

$$\begin{cases} \mathcal{L}[\psi(\eta)] = \int_0^\infty \psi(\eta) e^{-s\eta} d\eta = \Psi(s) \\ \mathcal{L}\left[\eta \frac{d\psi(\eta)}{d\eta}\right] = -s \frac{d\Psi(s)}{ds} - \Psi(s) \\ \mathcal{L}\left[\frac{d\psi(\eta)}{d\eta}\right] = -s\Psi(s) - \psi(0) \\ \mathcal{L}\left[\int_0^\eta \psi(\eta_1)\psi(\eta - \eta_1)d\eta_1\right] = \Psi(s)^2 \end{cases} \tag{30}$$

where $\mathcal{L}$ is the Laplace operator, $s$ is the Laplace transform parameter, $\Psi$ is the transformed dimensionless PSD, and $\psi(0) = 0$ for realistic kernel. Then the transformed governing equation of self-preserving size distribution can be rewritten as

$$-s \frac{d\Psi(s)}{ds} - \Psi(s) + \Psi(s)^2 = 0 \tag{31}$$

Its solution can be found as

$$\Psi(s) = \frac{1}{1+s} \tag{32}$$

Its inverse Laplace transformation is defined as

$$\mathcal{L}^{-1}[\Psi(s)] = \int_0^\infty \Psi(s) e^{s\eta} ds = \psi(\eta) \tag{33}$$

For classical Smoluchowski equation, the similarity solution can be found as

$$\psi(\eta) = e^{-\eta} \tag{34}$$

Based on the inverse similarity transformation, the solution of classical Smoluchowski equation can be written as

$$n(v,t) = \frac{M_0^2}{M_1} \exp\left[-\frac{M_0}{M_1} v\right] \tag{35}$$

Usually, the analytical similarity solution is independent of kernels and the boundary condition. therefore, the obtained solution may not match the experimental data or carry no actual physical significance. An improved method is to obtain the corrected similarity solution by iDNS using the present analytical similarity solution as initial condition, the physically realistic kernel and the corresponding boundary condition can be incorporated. A detailed description of AK-iDNS for the classical Smoluchowski coagulation equation can be found in our previous work (**Xie, 2023; Pan et al., 2024**).

**Diffusion coefficient based on particle moment**

The diffusion coefficient of particles in a fluid is a measure of how quickly particles spread out due to random motion. For nanoparticles, this can be influenced by their size, which can be characterized by the moments of the particle size distribution.

For spherical particles in the free molecule regime, the diffusion coefficient $D_n$ can be estimated using Einstein-Smoluchowski relation $D_n = k_B T / f$. To relate the diffusion coefficient to the moments of the particle size distribution, we can use an average volume or diameter. One common choice is the volume-weighted mean diffusion coefficient as

$$\overline{D_n} = \int_0^\infty D_n \frac{n(v,t)}{M_0} dv = \frac{1}{\frac{2}{3}\left(\frac{6}{\pi}\right)^{\frac{2}{3}}\rho\left(\frac{2\pi k_B T}{m}\right)^{\frac{1}{2}}\left(1+\frac{\pi \alpha_p}{8}\right)} \frac{1}{M_0} M_{-\frac{2}{3}} \quad (36)$$

The fractional particle moment can be expressed as (**Xie, 2016**)

$$M_k = \frac{M_1^k}{M_0^{k-1}}\left[1 + \frac{k(k-1)(M_C-1)}{2}\right] \quad (37)$$

where the dimensionless particle moment ($M_C$) is defined as

$$M_C = \frac{M_0 M_2}{M_1^2} \quad (38)$$

and

$$M_{-\frac{2}{3}} = \frac{4+5M_C}{9} M_0 v_a^{-\frac{2}{3}} \quad (39)$$

Then the average diffusion coefficient is

$$\overline{D_n} = \frac{1}{\frac{2}{3}\left(\frac{6}{\pi}\right)^{\frac{2}{3}}\rho\left(\frac{2\pi k_B T}{m}\right)^{\frac{1}{2}}\left(1+\frac{\pi \alpha_p}{8}\right)} \frac{4+5M_C}{9} v_a^{-\frac{2}{3}} \quad (40)$$

This relationship highlights how the diffusion coefficient depends on the statistical properties of the particle size distribution, providing a link between microscopic particle characteristics and macroscopic transport properties.

**Non-dimensionalization**

The governing equations are non-dimensionalized to simplify the treatment and analysis of the interaction between fluid and particle fields. It can be accomplished using the following relations:

$$t^* = \frac{t}{L/U}; x^* = \frac{x}{L}; y^* = \frac{y}{L}; u^* = \frac{u}{U}; v^* = \frac{v}{U}; p^* = \frac{p}{\rho U^2}; M_k^* = \frac{M_k}{M_{k0}}; v_a^* = \frac{v_a}{v_{a0}}; \quad (41)$$

In which the characteristic length $L$ is the initial momentum thickness of the mixing layer ($\theta$); the characteristic velocity $U$ is the velocity difference across the mixing layer ($2U_1$); $M_{k0}$ is the initial value of the $k$th moment. Substituting the relations given in above into the governing equations yields the familiar mass and momentum conservation equations (For brevity, the star symbol '*' is omitted thereafter):

$$\frac{\partial u}{\partial x} + \frac{\partial v}{\partial y} = 0$$
$$\frac{\partial u}{\partial t} + u\frac{\partial u}{\partial x} + v\frac{\partial u}{\partial y} = -\frac{\partial p}{\partial x} + \frac{1}{Re}\left(\frac{\partial^2 u}{\partial x^2} + \frac{\partial^2 u}{\partial y^2}\right) \quad (42)$$
$$\frac{\partial v}{\partial t} + u\frac{\partial v}{\partial x} + v\frac{\partial v}{\partial y} = -\frac{\partial p}{\partial y} + \frac{1}{Re}\left(\frac{\partial^2 v}{\partial x^2} + \frac{\partial^2 v}{\partial y^2}\right)$$

Where the Reynolds number is $Re = UL/\nu$. Similarly, the non-dimensionalized equations for the first three moment equations of the particle fields are given by (**Garrick et al., 2006; Settumba and Garrick, 2003; Xie et al., 2012**)

$$\begin{cases} \frac{\partial M_0}{\partial t} + u\frac{\partial M_0}{\partial x} + v\frac{\partial M_0}{\partial y} = \frac{1}{ReSc}\frac{\partial}{\partial x}\left(v_a^{-\frac{2}{3}}\frac{\partial M_0}{\partial x}\right) + \frac{1}{ReSc}\frac{\partial}{\partial y}\left(v_a^{-\frac{2}{3}}\frac{\partial M_0}{\partial y}\right) - \frac{1}{2}Dav_a^{\frac{1}{6}}M_0^2 \\ \frac{\partial M_1}{\partial t} + u\frac{\partial M_1}{\partial x} + v\frac{\partial M_1}{\partial y} = \frac{1}{ReSc}\frac{\partial}{\partial x}\left(v_a^{-\frac{2}{3}}\frac{\partial M_1}{\partial x}\right) + \frac{1}{ReSc}\frac{\partial}{\partial y}\left(v_a^{-\frac{2}{3}}\frac{\partial M_1}{\partial y}\right) \\ \frac{\partial M_2}{\partial t} + u\frac{\partial M_2}{\partial x} + v\frac{\partial M_2}{\partial y} = \frac{1}{ReSc}\frac{\partial}{\partial x}\left(v_a^{-\frac{2}{3}}\frac{\partial M_2}{\partial x}\right) + \frac{1}{ReSc}\frac{\partial}{\partial y}\left(v_a^{-\frac{2}{3}}\frac{\partial M_2}{\partial y}\right) + Dav_a^{\frac{1}{6}}M_1^2 \end{cases}$$
(43)

And the Schmidt number based on the particle moment is given as

$$Sc = \frac{v}{\kappa}v_{a0}^{\frac{2}{3}} \tag{44}$$

And the size independent diffusivity ($\kappa$) is

$$\kappa = \overline{D_n}v_a^{\frac{2}{3}} \tag{45}$$

The Damkohler number ($Da$) represents the ratio of the convective time scale to the coagulation time scale and is given by

$$Da = \frac{pM_{00}^2 v_{a0}^{\frac{1}{6}}}{U/L} \tag{46}$$

It is obvious that equations (45) are the system of partial difference equations and all the terms are denoted by the first three moments $M_0$, $M_1$ and $M_2$, and the system presents non closure problem. Under these conditions, the first three moments for describing particle dynamics are obtained through solving the systems of partial differential equations. Here, the deviation of equations (45) for particle fields does not involve any assumptions for the particle size distribution, and the final mathematical form is much simpler than the method of moment, QMOM, TEMOM, etc. Furthermore, only the first and second equation in the system of moment equations are coupled and necessary. However, in order to facilitate comparison with previous work, this study still adopts a three-equation model for the particle fields.

Initially, the lower stream is free of particles, while the upper stream is populated by nanoparticles. The initial $k$th moment for $k = 0,1,2$ are $M_{00} = 1, M_{10} = 1, M_{20} = 4/3$.

**Numerical algorithm for the evolution of fluid fields with time**

The general approach of the simulation of fluid flow is described as follows. While $u, v, p$ are the solutions to the Navier-Stokes equations, we denote the numerical approximations by capital letters. Assume we have the velocity field $U^i$ and $V^i$ at the $i^{th}$ time step (time $t$) and continue condition is satisfied. We find the solution at the $(i + 1)^{st}$ time step (time $t + \Delta t$) by the following three step approach (**Strang, 2007**).

The nonlinear terms are treated explicitly. This circumvents the solution of a nonlinear system, but introduces a CFL condition which limits the time step by a constant time the special resolution.

$$\begin{cases} \frac{U^* - U^i}{\Delta t} = -\left((U^i)^2\right)_x - (U^i V^i)_y \\ \frac{V^* - V^i}{\Delta t} = -(U^i V^i)_x - \left((V^i)^2\right)_y \end{cases} \tag{47}$$

The viscosity terms are treated implicitly. If they were treated explicitly, we would have a time step restriction proportional to the special discretization squared. We have no such limitation for the implicit treatment. The price to pay is two linear systems to be solved in each time step.

$$\begin{cases} \frac{U^{**}-U^*}{\Delta t} = \frac{1}{Re}\left(U^{**}_{xx} + U^{**}_{yy}\right) \\ \frac{V^{**}-V^*}{\Delta t} = \frac{1}{Re}\left(V^{**}_{xx} + V^{**}_{yy}\right) \end{cases} \quad (48)$$

We correct the intermediate velocity field $(U^{**}, V^{**})$ by the gradient of a pressure $P^{i+1}$ to enforce incompressibility.

$$\begin{cases} \frac{U^{i+1}-U^{**}}{\Delta t} = -\left(P^{i+1}\right)_x \\ \frac{V^{i+1}-V^{**}}{\Delta t} = -\left(P^{i+1}\right)_y \end{cases} \quad (49)$$

The pressure is denoted $P^{i+1}$, since it is only given implicitly. It is obtained by solving a linear system. In vector notation the correction equation read as

$$\begin{cases} \frac{\mathbf{U}^{i+1}-\mathbf{U}^i}{\Delta t} = -\nabla P^{i+1} \\ -\Delta P^{i+1} = -\frac{\nabla \mathbf{U}^i}{\Delta t} \end{cases} \quad (50)$$

Applying the divergence to both sides yield the linear system. Hence, the pressure correction step is: 1) Compute $F^i = \frac{\nabla \mathbf{U}^i}{\Delta t}$; 2) Solve Poisson equation $-\Delta P^{i+1} = -\frac{1}{\Delta t}F^i$; 3) Compute $\mathbf{G}^{i+1} = \Delta P^{i+1}$; 4) Updated velocity field $\mathbf{U}^{i+1} - \mathbf{U}^i = \Delta t \mathbf{G}^{i+1}$.

The question, which boundary conditions are appropriate for the Poisson equation for the pressure, is complicated. A standard approach is to prescribe homogeneous Neumann boundary condition for pressure where no-slip boundary conditions are prescribed for the velocity field.

**Compact finite difference scheme for the derivative in space**

Given the values of a function on a set of nodes the finite difference approximation to the derivative of the function is expressed as linear combination of the given function values. For simplicity consider a uniformly spaced mesh where the nodes are indexed by $i$. The independent variable at the nodes is $x_i = h(i-1)$ for $1 \le i \le N$ and the function values at the nodes $f_i = f(x_i)$ are given. The finite difference approximation $f_i'$ to the first derivative $df/dx(x_i)$ at the node $i$ depends on the function values at nodes near $i$. The compact finite difference schemes mimic the global dependence.

The generalizations are derived by writing approximation of the form (**Lele, 1992**):

$$\alpha f'_{i-1} + f'_i + \alpha f'_{i+1} = a \frac{(f_{i+1}-f_{i-1})}{2h} \quad (51)$$

The relations between the coefficient $\alpha$ and $a$ are derived by matching the Taylor series coefficients of various orders.

$$\begin{cases} f'_{i-1} = \frac{df_i}{dx} - \frac{d^2 f_i}{dx^2}h + \frac{1}{2}\frac{d^3 f_i}{dx^3}h^2 - \frac{1}{6}\frac{d^4 f_i}{dx^4}h^3 + O(h^4) \\ f'_{i+1} = \frac{df_i}{dx} + \frac{d^2 f_i}{dx^2}h + \frac{1}{2}\frac{d^3 f_i}{dx^3}h^2 + \frac{1}{6}\frac{d^4 f_i}{dx^4}h^3 + O(h^4) \\ f_{i-1} = f_i - \frac{df_i}{dx}h + \frac{1}{2}\frac{d^2 f_i}{dx^2}h^2 - \frac{1}{6}\frac{d^3 f_i}{dx^3}h^3 + \frac{1}{24}\frac{d^4 f_i}{dx^4}h^4 + O(h^5) \\ f_{i+1} = f_i + \frac{df_i}{dx}h + \frac{1}{2}\frac{d^2 f_i}{dx^2}h^2 + \frac{1}{6}\frac{d^3 f_i}{dx^3}h^3 + \frac{1}{24}\frac{d^4 f_i}{dx^4}h^4 + O(h^5) \end{cases}$$

And the constraints are

$$\begin{cases} 1 + 2\alpha = 2a \\ \alpha = \dfrac{a}{3} \end{cases} \tag{53}$$

And it can be obtained by:

$$\begin{cases} \alpha = \dfrac{1}{4} \\ a = \dfrac{3}{4} \end{cases} \tag{54}$$

The truncation error on the right-hand side for this scheme are $4/5!\,(3\alpha - 1)h^4 f^{(5)}$.

The first derivative for the non-period boundary condition at the left boundary $i = 1$ may be obtained from a relation of the form

$$f_1' + \alpha f_2' = \frac{1}{h}(a f_1 + b f_2 + c f_3 + d f_4) \tag{55}$$

Coupled to the relations written for the interior nodes. Requiring to be at least fourth-order accurate constraints the coefficients to, and the coefficient can be given by.

$$\alpha = 3;\ a = -\frac{17}{6};\ b = \frac{3}{2};\ c = \frac{3}{2};\ d = -\frac{1}{6} \tag{56}$$

The truncation error on the right-hand side for these boundary approximations are given by $6/5!\,h^4 f^{(5)}$.

Similarly, the first derivative at the right boundary $i = N$ can be expressed as

$$f_N' + \alpha f_{N-1}' = \frac{1}{h}(a f_N + b f_{N-1} + c f_{N-2} + d f_{N-3}) \tag{57}$$

And the coefficients are given by

$$\alpha = 3;\ a = \frac{17}{6};\ b = -\frac{3}{2};\ c = -\frac{3}{2};\ d = \frac{1}{6} \tag{58}$$

Analogously, the coefficient of the second-order differential compact finite difference scheme can be obtained, which are listed in Table 1.

Table 1. Coefficients of compact finite difference

|  | $\alpha$ | $a$ | $b$ | $c$ | $d$ |
| --- | --- | --- | --- | --- | --- |
| first-order derivative | | | | | |
| mid-point | 1/4 | 3/4 | | | |
| lower-end boundary | 3 | -17/6 | 3/2 | 3/2 | -1/6 |
| upper-end boundary | 3 | 17/6 | -3/2 | -3/2 | 1/6 |
| second-order derivative | | | | | |
| mid-point | 1/10 | 6/5 | | | |
| lower-end boundary | 11 | 13 | -27 | 15 | -1 |
| upper-end boundary | 11 | 13 | -27 | 15 | -1 |

Table 2. The comparison of computational time for different grids and method

| grids | Time (s) | | |
|---|---|---|---|
| | DNS | DNS+TEMOM | DNS+AK-iDNS |
| 64×64 | 70.1905 | 169.6806 | 124.462974 |
| 128×128 | 147.5984 | 463.8003 | 230.230999 |
| 256×256 | 504.1866 | 1665.649 | 692.841514 |

## Results and Discussion

**The effect of resolution on the fluid flow**

The resolution of a fluid flow simulation, defined by the fineness of computational grid or the number of discrete points used to represent the flow domain, has a significant impact on the accuracy and reliability of the results. When simulating fluid flows, particularly shear flows in a mixing layer, the grid resolution plays a crucial role in accurately capturing the dynamics of vortices and other flow structures. Usually, DNS requires resolving all scales of motion down to the Kolmogorov scale, and the number of grid points ($N$) needed in each dimension scales with the Reynolds number as $N \sim Re^{3/4}$. **Figure 2** shows the vorticity distribution during the rolling and pairing stages of vortices at a Reynolds number of 200. As the number of grids increases, the scale range of the vortices (velocity gradient) becomes smaller and rotational speed of vortices slows down. The reason for this phenomenon may lie in keeping the Reynolds number constant implies that the ratio of inertial forces to viscous forces remains the same, but the resolution of the simulation can still impact the observed flow characteristics. Coarse grids can introduce numerical diffusion, which artificially smooths out velocity gradients and can lead to an overestimation of vortex rotation speeds. Increasing the grid resolution reduces numerical diffusion, leading to more accurate representation of the flow field and a more realistic decreases in vortex rotation speeds.

Furthermore, this phenomenon can also be analyzed from the perspective of velocity distribution as shown in Figure 3. According to the definition of vorticity,

$$\Omega = \frac{\partial U}{\partial y} - \frac{\partial V}{\partial x} \tag{59}$$

From **Figure 3b and 3c**, it can be seen that during the rolling and pairing stage of the vortex, the velocity near the center of the vortex is basically linear with the coordinates, and the velocity gradient is basically consistent under different grid resolution conditions, i.e., the magnitude of the vorticity is consistent. But as the grid resolution increases, the range of vortices becomes smaller, and the relative average flow velocity of the vortex decreases, resulting in slower rotation speed. Therefore, higher grid resolution allows for the resolution of smaller scales of motion, and fine resolution is necessary to accurately capture the interfaces between different fluids in the mixing layer, where steep velocity gradient occur. However, higher resolution significantly increases the computational cost in terms of memory and processing time based on CFL conditions to avoid numerical diffusion or dispersion. The number of grid points and the time steps required for stability both increases, leading to longer simulation times and greater resource requirements. The comparison of computational time for different grids is listed in Table 2. The simulation conditions are that the time step is 0.01, the total dimensionless time is 0-100 seconds, the computing platform is the MacBook Pro with 8 core i9 2.3GHz Intel CPU and 16 GB 2667 Hz DDR4 memory, the software is MATLAB R2023a. As the grid resolution increases, the computational cost shows a sharp exponential increase. Therefore, balancing resolution with available computational resources is a key consideration in practical simulation.

**Figure 4** shows the vorticity distribution during the merging stage of vortices under the condition of Reynolds number of 200, while **Figure 5** presents the velocity distribution in the central horizontal and vertical directions. The impact of grid resolution on vortices is minimal, and the velocity distribution is generally consistent with each other. This means that in the fully developed stage of the mixing layer, the higher grids resolution does not necessarily lead to higher vortex resolution. In response to the above situation, the adaptive mesh refinement techniques can help optimize this trade-off by providing high resolution where needed while conserving resources elsewhere. Balancing resolution with available resources and the specific requirements of the present mixing layer simulations, this article mainly uses a grid resolution of 256×256 for effective and efficient computation.

**The evolution of vortices with time**

The evolution of a temporal mixing layer is a fundamental phenomenon in fluid dynamics, driven by several key mechanisms and factors. A temporal mixing layer forms when two parallel streams of fluid with different velocities interact, leading to the development of complex flow structures over time. The primary driver of the evolution of a temporal mixing layer is shear instability. The velocity difference between the two fluid streams creates a shear layer, which becomes unstable and amplifies small perturbations. This instability leads to the formation of vortices and coherent structures. The linear stability theory of temporal mixing layer is shown in the **Appendix I. Figure 6** shows the distribution of eigenvalues based on linear stability theory. For temporal mixing layer, the eigenvalue with maximum imaginary part is $\omega_i = 0.3418$; for Couette flow, the eigenvalue with maximum imaginary part is $\omega_i = 0$. From the perspective of linear stability theory, the temporal mixing layer is unstable. While Couette flow is linearly stable for all Reynolds numbers, meaning that infinitesimally small perturbations do not grow exponentially. However, the flow can transition to turbulence through nonlinear mechanisms and finite-amplitude perturbations. The study of nonlinear stability theory is beyond the scope of this article and will not be discussed here.

As the shear layer evolves, vortices form and grow due to the Kelvin-Helmholtz instability. These vortices interact, merge, and break down into smaller structures, contributing to the development of coherent structure and enhancing mixing between the fluid streams. The evolving vortices and coherent structures entrain fluid from the surrounding streams, increasing the mixing of momentum, energy, and scalar quantities (such as temperature or concentration). This mixing process is crucial for the homogenization of the fluid properties across the layer. **Figure 7** shows the evolution of the main vortex in the mixing layer over time from the beginning to the merging.

External factors such as pressure gradients, boundary conditions and external forcing can influence the evolution of the mixing layer. These factors can either stabilize or destabilize the shear layer, affecting the growth and dynamics of the vortices. Viscous effects play an important role in the evolution of the mixing layer by dissipating kinetic energy at small scales. This dissipation limits the growth of turbulence and contributes to the eventual decay of the mixing layer if no additional energy is supplied. In this article, there is no introduction of external forces in the temporal mixing layer, and the vortex gradually decreases under the influence of boundary conditions and viscous dissipation. **Figure 8** shows the gradual degeneration of the mixing layer into the Couette flow due to the viscous dissipation mechanism after vortex merging. **Figure 9** shows the evolution of velocity in the central horizontal and vertical directions over time under the condition of Reynolds number of 200. It can be seen that the distribution of horizontal velocity along the y-coordinate ultimately exhibits linear profile characteristic of Couette flow.

In summary, the evolution of a temporal mixing layer is driven by shear instability, vortex formation and interaction, entrainment and mixing, nonlinear effects, external forcing and viscous dissipation. The evolution of a temporal mixing layer can degenerate into the Couette flow under conditions that suppress shear instabilities, such as strong viscous damping, low Reynolds number and specific boundary conditions. Understanding these mechanisms is essential for predicting and controlling the behavior of mixing layers in various engineering and natural system.

**The effect of Reynolds number on the evolution of vortices**

The Reynolds number ($Re$) is a dimensionless quantity that characterizes the relative importance of inertial forces to viscous forces in a fluid flow. It plays a crucial role in determine the behavior and evolution of a temporal mixing layer. At low Reynolds numbers, the flow is dominated by viscous forces, and the mixing layer remains relatively stable. Small perturbation in the flow tend to decay, and the layer evolves smoothly without significant instabilities. The growth rate of the mixing layer is slow, and the velocity profile remains relatively linear with a gradual transition between the two streams. At high Reynolds numbers, the flow becomes fully turbulent. The mixing layer is characterized by a wide range of scales of motion, from large energy-containing eddies to small dissipative scales. Turbulence significantly enhances the mixing and entrainment processes. The energy cascade process becomes prominent.

This article mainly focuses on temporal mixing layer at intermediate Reynolds number. In the transitional regime, as the Reynolds number increases, inertial forces become more significant, leading to the onset of shear instabilities. These instabilities can grow and lead to the formation of vortices and coherent structure within the mixing layer. The Kelvin-Helmholtz instability is a common mechanism that drives the formation of vortices in the mixing layer. These vortices enhance mixing and entrainment between two streams. The growth rate of the mixing layer increases with the Reynolds number. At higher Reynolds numbers, the mixing layer thickens more rapidly due to the enhanced mixing and entrainment as shown in **Figure 10**. The result can also be obtained through the location of the maximum velocity in the velocity distribution as listed in **Figure 11**.

**The comparison of the distribution of particle moments between TEMOM and AK-iDNS**

In this article, both the average kernel method (AK-iDNS) and the TEMOM are techniques used to solve PBE that describe the evolution of particle size distribution due to processes like coagulation and mixing. The dimensionless time for the system to achieve self-preserving size distribution for zero-dimensional particle coagulation is about 10 seconds. **Figure 12 and Figure 13** are the comparison of the AK and TEMOM in the temporal mixing layer at 20 seconds.

In fluid dynamics, both the distribution of particle moments and the distribution of vortices exhibit certain similarities and scaling behavior in their statistical and spatial characteristics, both average kernel method and TEMOM can capture the similarity of statistical properties as shown in **Figure 12** and **Figure 13**. These similarities arise from the underlying physical processes (such as rolling, pairing and merging, etc.) that govern their evolution and interactions. The moments of the particle size distribution provide statistical measures of the distribution's shape and spread. For example, the zeroth moment represents the total number of the particles, the first moment represents the total mass or volume, and higher order moments provide information about the distribution's variance and skewness. The evolution of particle moment is influenced by processes such as advection, diffusion and coagulation. These processes change the particle size distribution over time. The distribution of vortices in the mixing layer can also be described statically. Measures such as the number of vortices, their circulation strength and their spatial distribution provide insights into the flow characteristics. Similarly, the evolution of vortices is influenced by processes such as vortex rolling, pairing and merging. These processes change the distribution of vortices in the flow. In addition, Particle in the mixing layer can exhibit clustering due to the interaction with coherent structures such as vortices. This clustering affects the local particle concentration and size distribution.

Careful comparison between **Figure 12 and Figure 13** reveals that the distribution of vortices and the first order particle moment are consistent in both average kernel method and TEMOM. However, there is a significant difference in the distribution of zeroth and second order particle moments, which is mainly reflected in their amplitude. The reason is that the average kernel method simplifies the PBE by approximating the coagulation kernel with an average value. This reduces the complexity of the integral terms in the PBE. It is computationally efficient, making it suitable for real-time applications and large-scale simulations. The TEMOM approximates the moments of particle size distribution using a Taylor series expansion. This method transforms the PBE into set of ordinary different equations (ODEs) for the moments. TEMOM method provides a more accurate representation of the particle size distribution, especially for system with complex kernels. The TEMOM is more complex and computationally intensive compared to the average kernel method as shown in **Table 2**.

The different amplitudes of the distribution of particle moments affect the statistical characteristics of the particle size distribution. **Figure 14** show the distribution of algebraic mean volume and dimensionless particle moment. The results with the average kernel method have smaller average particle size and variance. Through the contour maps, it can also be seen that the relative distribution of zeroth and second order particle moments in space are consistent with each other, which means that both AK-iDNS and TEMOM have the same scaling law for particle coagulation. Therefore, the choice between the two methods depends on the specific requirements of the application, including the desired accuracy and computational resources.

**The evolution of particle moments over time**

Coagulation is a process where particles collide and stick together, forming larger particles, which affects the particle size distribution and, consequently, the moments of the particle size distribution. In a temporal mixing layer, the evolution of particle moments due to coagulation can be significantly influenced by the flow dynamics. **Figure 15** show the evolution of particle moments with time at different point along the central horizontal and vertical line in the flow field under the conditions $Re = 200, Sc = 1, Da = 1$. Overall, the evolution of particles along the central horizontal line can be divided into three stages as shown in the left of **Figure 15**. In the initial stage, the coherent structures are not generated, and the evolution of particles is basically consistent with the 0-dimensional problems, i.e., the zeroth order particle moment decreases exponentially over time, the first order particle moment almost remains constant, and the second particle moment increases exponentially over time. During the rolling, pairing and merging stages of vortices, the evolution of particle moments exhibits significant fluctuations due to the effect of flow dynamics, such as advection, mixing and diffusion, etc. In the dissipation stage of the vortex, the effect of the flow dynamics gradually disappears, and the evolution of particles returns to the 0-dimensional problem.

Along the central vertical line in the flow field, the evolution of particles at different points varies greatly as shown in the right of **Figure 15**. According to the distribution of particles in space. They can be roughly divided into three regions. At the top of the flow field, the influence of vortex structure is minimal, and the evolution of particles is almost consistent with the 0-dimensional problem. At the bottom of the flow field, there are no particles in the initial stage. Due to the advection and diffusion of flow dynamics, the particle number density and volume density continue to increase, and the dispersity of particle size spectrum is also increasing continually. Due to the evolution of vortices, the evolution of particle moments also exhibits certain fluctuations. In the influence area of vortices, the evolution of particle presents a more complex morphology. In the upper part of the flow field, similar to the evolution of particles along the central horizonal line, the evolution of particles can be divided into three stages. In the lower part of the flow field, due to the initial absence of particles, the number density, volume density together with the dispersity show an increasing trend for the entrainment and mixing effect of the flow dynamics in the initial stage. The evolution of particle in the latter two stages is similar to that in the upper part.

**Figure 16** show the evolution of particle algebraic mean volume ($v_a$) and dimensionless moments ($M_C$) at different points along the central horizontal and vertical line with time. At different points, the algebraic mean volume exhibits exponential growth over time. However, the evolution of particles in the central horizontal line follows a consistent asymptotic growth rate. Despite the fluctuations of vortices, the evolution curves basically coincide with each other at long time. In contrast, the asymptotic growth rate in the central vertical line is almost the same, but there are some differences in the absolute values of the evolution curves. Analogously, the dimensionless particle moments at different points along the central horizontal line tend towards a constant at long time. While the evolution of dimensionless moments ($M_C$) at different points along the central vertical line tends towards different constants. The difference mainly comes from the transport and mixing effects of flow dynamics on particle coagulation. Of course, the diffusion effect of particles cannot be ignored, which will discussed in later parts.

**The effect of Damkohler number**

The Damkohler number ($Da$) is a dimensionless quantity that characterizes the relative importance of particle dynamics to transport processes (such as advection, diffusion, etc.) in a system. In the context of particle laden flows, the Damkohler number can significantly influence the evolution of particle moments, especially when coagulation processes are involved. When Damkohler number is much less than 1 ($Da \ll 1$), the transport processes dominate over the coagulation. The evolution of particle moments is primarily governed by the mixing and diffusion of particles. Coagulation occurs slowly, and the particle size distribution changes gradually over time. When Damkohler number is much greater than 1 ($Da \gg 1$), the coagulation process dominates over the transport processes. In this regime, coagulation occurs rapidly, leading to significant changes in the particle size distribution. Particles collide and coalesce quickly, forming larger particles. When Damkohler number is around 1 ($Da \sim 1$), the coagulation and transport processes are balanced. In this regime both coagulation and mixing influence the evolution of particle moments. The particle size distribution evolves due to a combination of coagulation and transport effects as shown in **Figure 17**.

The Damkohler number has almost no effect on the distribution of the particle volume concentration, which can also be seen from the source term of the first order moment equation. due to the volume conservation of coagulation, the source term is 0. In addition, the spatial distribution of particle moments is similar to that of vortices, and Damkohler number mainly affects the mean and amplitude of zeroth and second order moments.

**The effect of Schmidt number**

The Schmidt number $Sc$ is a dimensionless quantity that characterizes the ratio of momentum diffusivity (viscosity) to mass diffusivity (diffusion) in a fluid. It plays a significant role in determining the behavior of particles in a fluid flow, particularly in processes involving diffusion and mixing. When $Sc$ is much less 1, the mass diffusivity is much greater than the kinematic viscosity. This means that particles diffuse rapidly compared to the rate at which momentum diffuses. In this regime, the rapid diffusion of particles leads to a more uniform distribution of particles in the fluid. Coagulation and other particle interactions are influenced by the enhanced mixing and dispersion. When $Sc$ is much greater than 1, the mass diffusivity is much less than the kinematic viscosity. This means that particles diffuse slowly compared to the rate at which momentum diffuses. In this regime, the slow diffusion of particles leads to higher local concentrations of particles, increasing the likelihood of collisions and coagulation. When $Sc$ is around 1, the mass diffusivity and kinematic viscosity are comparable. This means that the rate of particle diffusion and momentum diffusion are balanced. In this regime, both diffusion and coagulation processes significantly influence the evolution of particle moments. The particle size distribution evolves due to a combination of mixing and particle interactions. The moments evolve in a more complex manner as shown in **Figure 18**, with both lower order and higher order moments changes due to the interplay between diffusion and coagulation. The influence of Schmidt number revolves around a baseline vibration, the larger the Schmidt number, the greater the fluctuation amplitude.

In addition, the reciprocal of the product of Schmidt number and Reynolds number in the diffusion term of particle evolution equation. therefore, the effect of Schmidt number on the evolution of particles is equivalent to the effect of Reynolds number.

**The effect of advection on the particle size distribution**

Advection is the transport of particles and vortices by the bulk motion of the fluid, this means that particles move with the flow, maintaining their relative positions and structures as they are carried by the fluid. This is the main reason for the similarity of the distribution between the particle moment and the vortices. Advection, interacting with shear layers and coherent structures, contribute to processes such as vortex stretching, tilting, merging and splitting. These interactions play a crucial role in the energy cascade and overall dynamics of flows as shown in **Figure 7**. While advection primarily affects the spatial distribution of particles, it can indirectly influence the particle size distribution under certain conditions.

Advection moves particles along with the fluid flow, redistributing them spatially within the flow domain. This process does not direct change the size of individual particles but affects where particles of different size are located. By redistributing particles, advection can bring particles of different sizes into closer proximity, potentially increasing the collision rate. This can enhance coagulation leading to changes in the particle size distribution.

Advection can create or modify concentration gradients of particles within the flow, which can influence local particle interactions and processes such as coagulation. When combined with coagulation, advection can significantly impact of the evolution of the particle size distribution. The advection term in the particle PBE can influence the spatial distribution of particles, which in turn affects the local coagulation rates. Advection can work in conjunction with diffusion to spread particles throughout the flow domain. The combined effect of advection and diffusion can lead to a more uniform distribution of particles, affecting the overall size distribution. Overall, in the mixing layer, advection can particles from regions of high concentration to regions of low concentration. This redistribution can lead to increased local particle concentrations and higher collision rates, resulting in changes to the particle size distribution over time.

**The asymptotic particle size distribution**

Under asymptotic conditions, the distribution of particle moments is shown in **Figure 19.** Due to the limitation in computational resources, this article only calculates up to the dimensionless time $t = 1000\ s$. In the later stage, the influence of the flow field on the evolution of particles becomes smaller and smaller, and longer simulations seems unnecessary. The results reveal that the particle distribution in the horizontal direction is uniform, or the gradient of particle moments is zero, i.e., $\partial n/\partial x = 0$. In the vertical direction, the distribution of particle moments is basically linear, which means that the second derivative of particle moments in the vertical direction is zero. Then the diffusion term in the PBE can be eliminated.

Together with the asymptotic distribution of velocities in Figure 9 ($V = 0$), the advection term can also be eliminated. And the particle population balance equation degenerates into the Smoluchowski coagulation equation, i.e.,

$$\frac{\partial n}{\partial t} = \left[\frac{\partial n}{\partial t}\right]_{coag} \tag{3}$$

Therefore, the particle size distribution under asymptotic conditions is consistent with the 0-dimenaional problem, which have been discussed in our previous work in details, and it will not be discussed here.

## Conclusions

A temporal mixing layer is a fundamental flow configuration where two parallel streams of fluid with different velocities interact, creating a shear layer that evolves over time. When nanoparticles are present in this flow, their evolution is influenced by various physical processes and the unique characteristics of the mixing layer. At the start, nanoparticles are typically distributed non-uniformly across the mixing layer, concentrated in the upper of the fluid streams. As the mixing layer develops, large-scale vortices form due to Kelvin-Helmholtz instabilities, which is captured through direct numerical simulation and linear stability analysis. These vortices enhance mixing, causing nanoparticles to be transported across the layer. Advection by the mean flow and large eddies redistributes particles throughout the mixing region, this enhances mixing and leads to a more uniform distribution over time. Brownian motion causes nanoparticle to diffuse, gradually spreading them across the mixing layer. The diffusion rate depends on particle size, with smaller particles diffusing more rapidly. To simplify the calculation, a moment based average diffusion coefficient model is proposed. Increased particle interactions can lead to coagulation, larger particles may form as smaller nanoparticles collide and stick together, the AK-iDNS framework is used to approximate the Smoluchowski coagulation equation. The results based on average kernel method and TEMOM are basically consistent, and both have the same scaling growth rate. However, the average kernel method has much higher computational efficiency. Strong velocity gradients in the mixing layer can cause shear-induced migration of particles, which can lead to non-uniform spatial distributions and preferential concentration in certain regions of the flow. Over time, a more homogeneous distribution is approached in the horizontal direction. The particles exhibit similar asymptotic behavior at that of 0-dimensional problem. However, due to the non-uniform spatial distribution of particles in the vertical direction, there are some differences in the distribution of the algebraic mean volume and dimensionless particle moment. But the evolution of particle moments has almost the same scaling growth rate.

In short, the evolution of nanoparticles in a temporal mixing layer is a complex process involving advection, diffusion, dispersion, and particle interactions. Initially non-uniform distributions become more homogeneous over time due to mixing processes. The interplay between particle characteristics and flow dynamics lead to unique spatial and temporal patterns in the particle distribution. Understanding the evolution is crucial for application in nanotechnology, environmental science and industrial processes involving nanoparticle-laden flow.

# Acknowledgements

This work was funded by the National Natural Science Foundation of China with grant number 11972169.

# Nomenclature

- $\rho$     fluid density
- $p$     pressure
- $p'$     disturbances of pressure
- $\nu$     kinematic viscosity of fluid
- $u$     velocity component in the $x$ direction
- $v$     velocity component in the $y$ direction
- $U$     base flow in in the $x$ direction or the characteristic velocity
- $V$     base flow in in the $y$ direction
- $u'$     disturbances in in the $x$ direction
- $v'$     disturbances in in the $y$ direction
- $U_1$     far field velocity at the upper flow
- $U_2$     far field velocity at the lower flow
- $x$     the Cartesian coordinate in the stream-wise direction
- $y$     the Cartesian coordinate in the transverse direction
- $\theta$     the initial momentum thickness

| | |
|---|---|
| $n$ | the number density of particles |
| $v$ | particle volume |
| $v_a$ | the algebraic mean volume |
| $t$ | time |
| $\beta$ | the collision kernel of coagulation |
| $\bar{\beta}$ | the average collision kernel |
| $k_B$ | the Boltzmann's constant |
| $T$ | the fluid temperature |
| $\rho_p$ | the particle density |
| $D_n$ | the particle diffusion coefficient |
| $f$ | the friction coefficient |
| $m$ | the molecular mass of fluid molecules. |
| $\alpha_p$ | the accommodation coefficient, |
| $d_p$ | the diameter of particles |
| $Da$ | the Damkohler number |
| $Re$ | the Reynolds number |
| $St$ | the Stokes number |
| $Sc$ | the Schmidt number based on the particle moment |
| $L$ | the characteristic length |
| $\alpha$ | the scaler factor or coefficient |
| $\alpha_p$ | the accommodation coefficient |
| $p$ | the proportional factor |
| $q$ | the power index |
| $M_0$ | total particle number concentration |
| $M_1$ | the particle volume concentration |
| $M_2$ | the dispersity of particle size distribution |
| $M_k$ | $k$th order moment of the volume-based particle size distribution |
| $M_{k0}$ | the initial value of $k$th order moment |
| $\eta$ | the dimensionless particle volume, |
| $\psi$ | the dimensionless particle size distribution |
| $\mathcal{L}$ | the Laplace operator, |
| $s$ | the Laplace transform parameter, |
| $\Psi$ | the transformed dimensionless particle size distribution |
| $A$ | the total collision frequency |
| $C$ | the gain term |
| $g$ | the loss term |
| $\varepsilon$ | error limits |
| $k$ | the spatial wavenumber in $x$ direction |
| $\omega$ | the temporal pulsation |
| $a, b, c, d$ | the coefficient |
| $\phi$ | the stream function |
| AK | average kernel |
| DNS | direct numerical simulation |
| iDNS | iterative direct numerical simulation |

PSD     particle size distribution

# Appendix I: Linear stability theory for parallel flow

The theory of stability of laminar flows decomposes the motion into a mean flow and a disturbance superimposed on it (Schmid and Henningson, 2001; Drain, 2002). Let the mean flow, which may be regarded as steady, be described by its Cartesian velocity component $U$, $V$ and pressure $P$. The corresponding quantities for the non-steady disturbance will be denoted by $u'$, $v'$ and $p'$, respectively.

$$\begin{cases} u = U + u' \\ v = V + v' \\ p = P + p' \end{cases} \quad (AI.1)$$

The Navier-Stokes equations linearized about the base flow profile $U$ in two-dimensions is

$$\begin{cases} \frac{\partial u'}{\partial x} + \frac{\partial v'}{\partial y} = 0 \\ \frac{\partial u'}{\partial t} + U\frac{\partial u'}{\partial x} + v'\frac{dU}{dy} = -\frac{\partial p'}{\partial x} + \frac{1}{Re}\Delta u' \\ \frac{\partial u'}{\partial t} + U\frac{\partial v'}{\partial x} = -\frac{\partial p'}{\partial y} + \frac{1}{Re}\Delta v' \end{cases} \quad (AI.2)$$

where $\Delta$ is the Laplacian. Rewriting this equation in operator form as

$$\left[-\begin{pmatrix} \partial_t & 0 & 0 \\ 0 & \partial_t & 0 \\ 0 & 0 & 0 \end{pmatrix} + \begin{pmatrix} -U\partial_x + \Delta/Re & dU/dy & -\partial_x \\ 0 & -U\partial_x + \Delta/Re & -\partial_y \\ \partial_x & \partial_y & 0 \end{pmatrix}\right] \begin{pmatrix} u' \\ v' \\ p' \end{pmatrix} = 0 \quad (AI.3)$$

where $\partial_{\{x,y,t\}}$ denote the partial derivative operators. We now consider normal modes, that is we assume an oscillating behavior in $x$ and time of the flow solution

$$\begin{cases} u' = u'(y)\exp[i(kx - \omega t)] \\ v' = v'(y)\exp[i(kx - \omega t)] \\ p' = p'(y)\exp[i(kx - \omega t)] \end{cases} \quad (AI.4)$$

where $k$ is the spatial wavenumber in $x$ direction and $\omega$ is the temporal pulsation. The exponential structure allows the solution to oscillate and grow/decay in space and time, depending on the real and imaginary parts of $k$ and $\omega$. In the temporal analysis, the solution grows/decay and oscillates in time but only oscillate in space: $k \in R$ is given, and one obtains $\omega$ from the dynamic equations.

As it has already been assumed that the perturbation is two-dimensional, it is possible to introduce a stream function $\phi(x, y, t)$ thus integrating the equation of continuity. The stream function representing a single oscillation of the disturbance is assumed to be of the form

$$\phi(x, y, t) = \phi(y)\exp[i(kx - \omega t)] \quad (AI.5)$$

From the stream function, it is possible to obtain the components of the perturbation velocity

$$\begin{cases} u' = \frac{d\phi}{dy} \\ v' = -ik\phi \end{cases} \quad (AI.6)$$

After the elimination of pressure, the following fourth-order ordinary differential equation for the disturbance can be obtained as

$$\frac{1}{ikRe}\left(\frac{d^4}{dy^4} - 2k^2\frac{d^2}{dy^2} + k^4\right)\phi + \frac{d^2U}{dy^2}\phi - U\left(\frac{d^2}{dy^2} - k^2\right)\phi = -\frac{\omega}{k}\left(\frac{d^2}{dy^2} - k^2\right)\phi \quad (AI.7)$$

It is commonly referred to as the Orr-Sommerfeld equation. by way of example, the boundary conditions for the temporal mixing layer demand that the components of the perturbation velocity must vanish at a large distance from the interface (free stream).

$$u' = 0, and \ v' = 0 \ at \ y = \pm\infty \tag{AI.8}$$

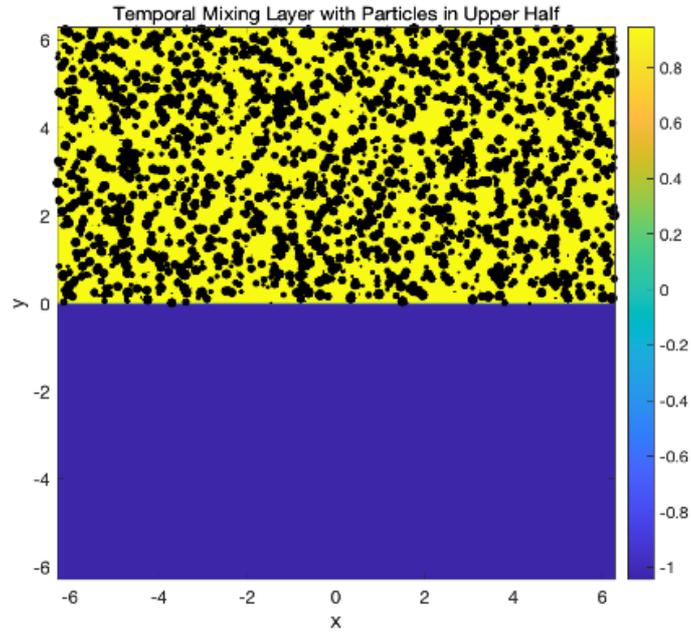

Figure 1. The configuration of nanoparticles in the Temporal mixing layer.

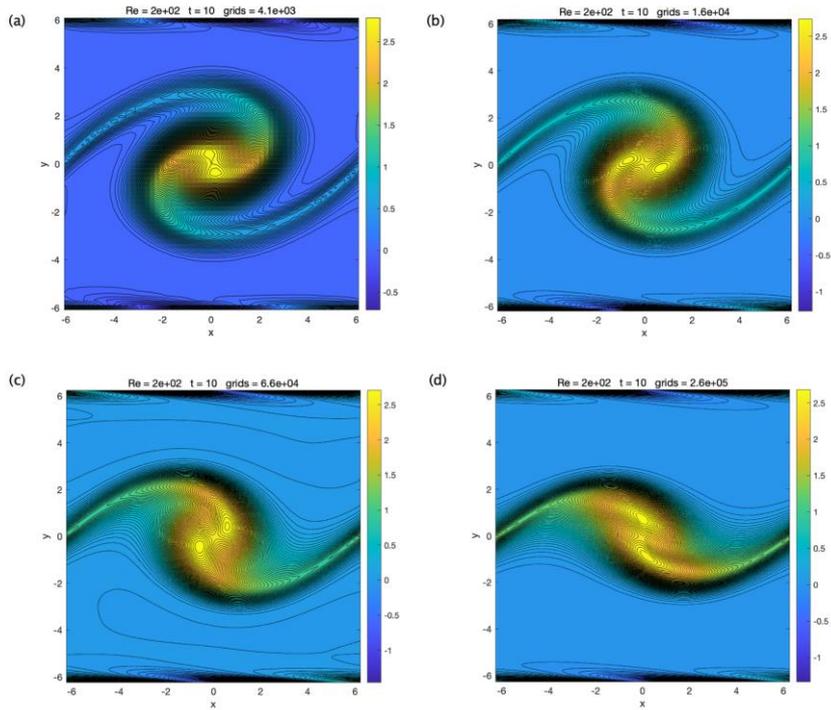

Figure 2. The vorticity distribution during the rolling and pairing stages of vortices at a Reynolds number of 200 (t = 10s) for different grids resolutions. a) 64×64 grids; b) 128×128 grids; c) 256×256 grids; d) 512×512 grids.

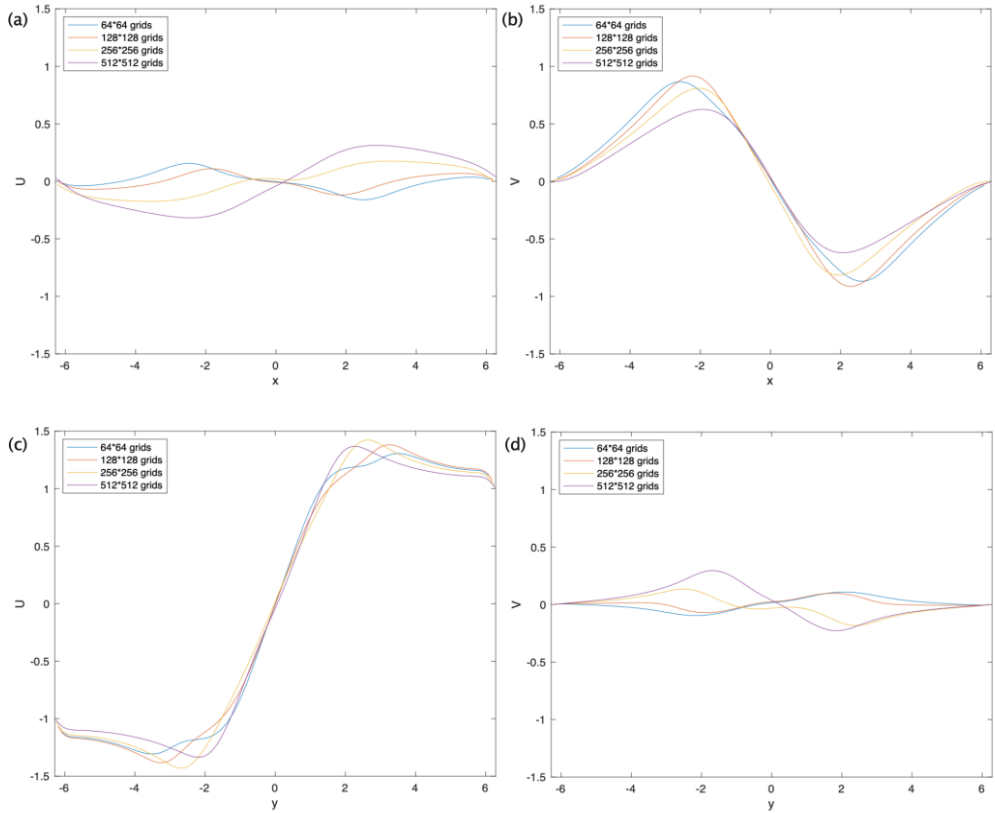

Figure 3. The velocity distribution in the central horizontal and vertical directions during the rolling and pairing stages of vortices at a Reynolds number of 200 (t = 10s). a) the relationship U-x; b) the relationship of V-x; c) the relationship of U-y; the relationship of V-y

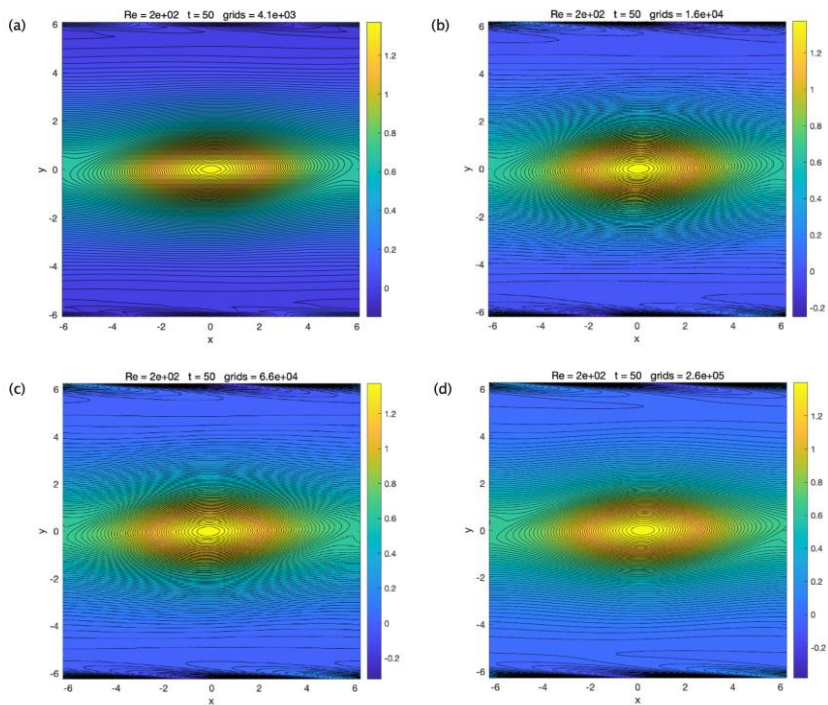

Figure 4. the effect of grid resolution on the fluid flow during the merging stage of vortices under the condition of Reynolds number of 200 (t=50s). a) 64×64 grids; b) 128×128 grids; c)

256×256 grids; d) 512×512 grids.

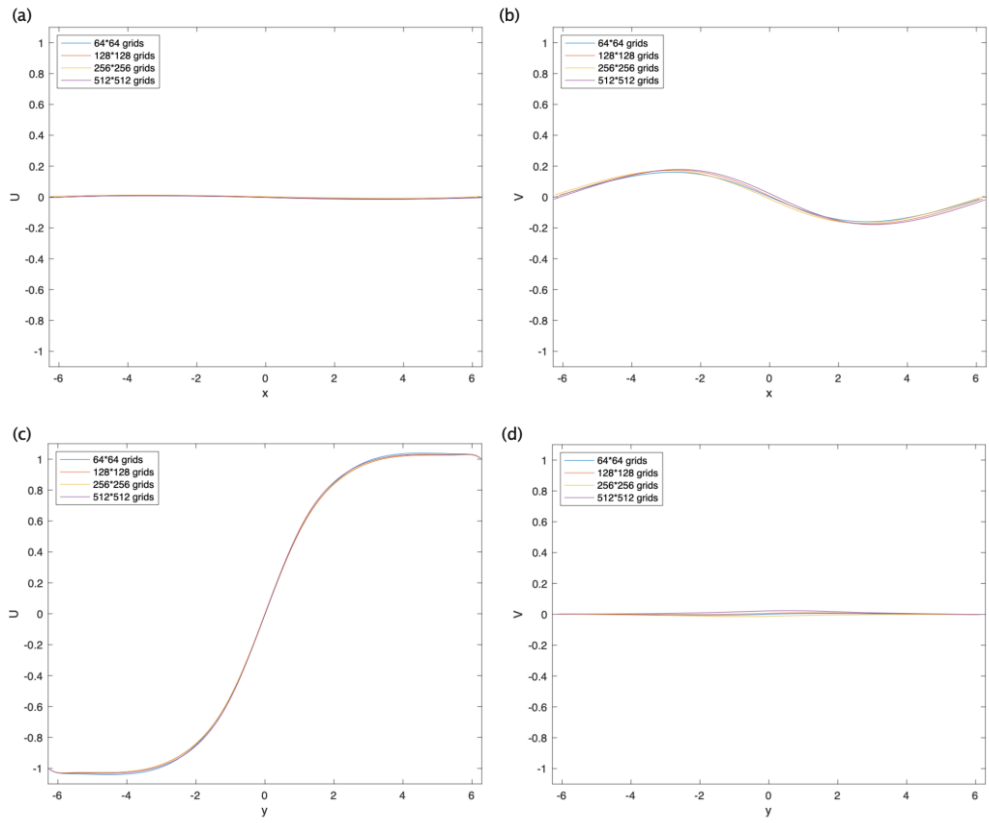

Figure 5. the velocity distribution in the central horizontal and vertical directions during the merging stage of vortices under the condition of Reynolds number of 200 (t=50s). a) the relationship U-x; b) the relationship of V-x; c) the relationship of U-y; the relationship of V-y.

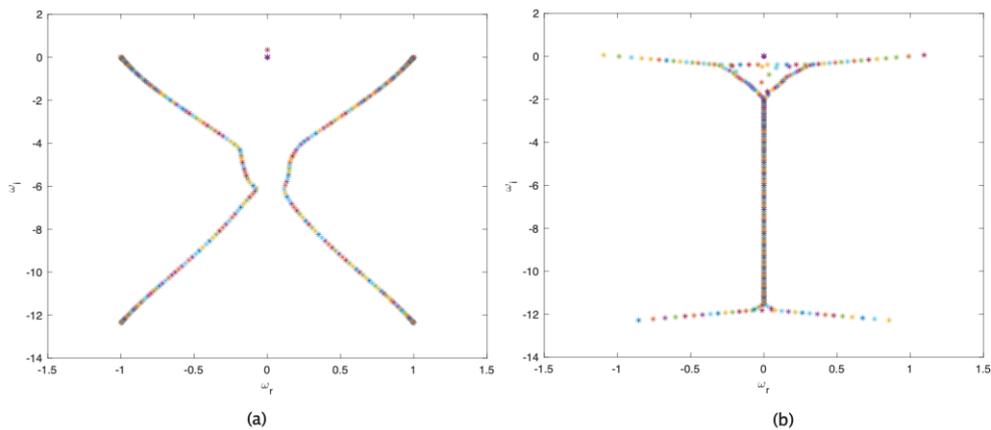

Figure 6. The eigenvalue based on linear stability theory for parallel flow. a) the temporal mixing layer flow; b) the Couette flow.

256×256 grids; d) 512×512 grids.

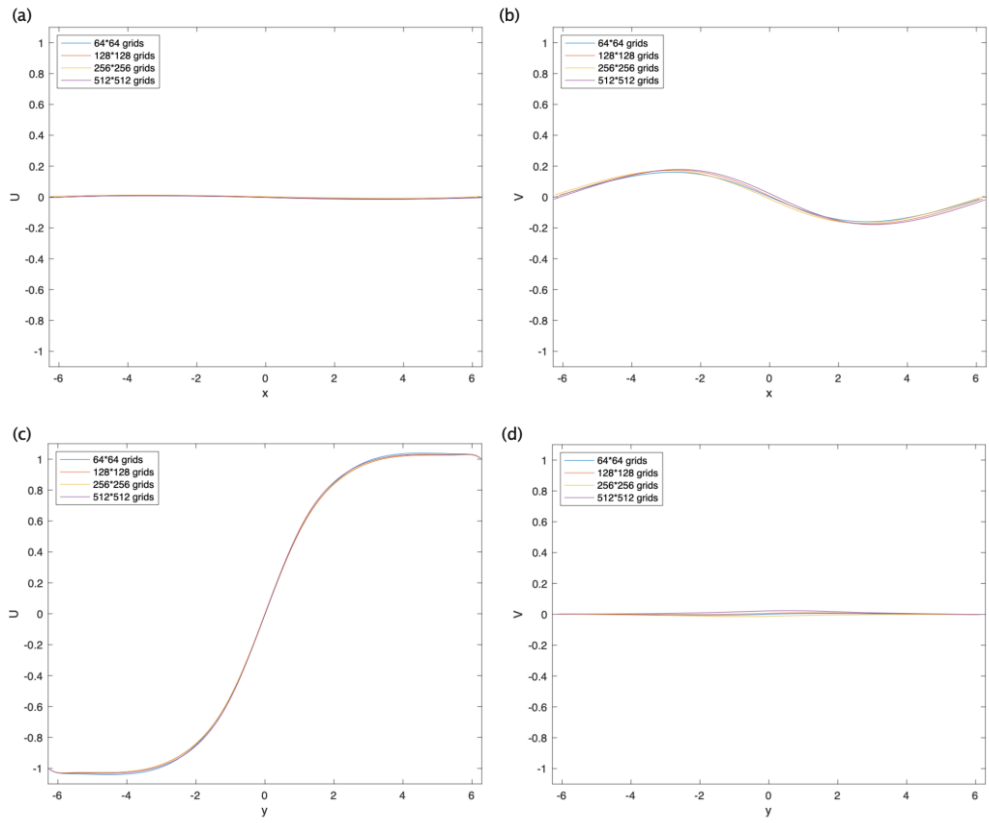

Figure 5. the velocity distribution in the central horizontal and vertical directions during the merging stage of vortices under the condition of Reynolds number of 200 (t=50s). a) the relationship U-x; b) the relationship of V-x; c) the relationship of U-y; the relationship of V-y.

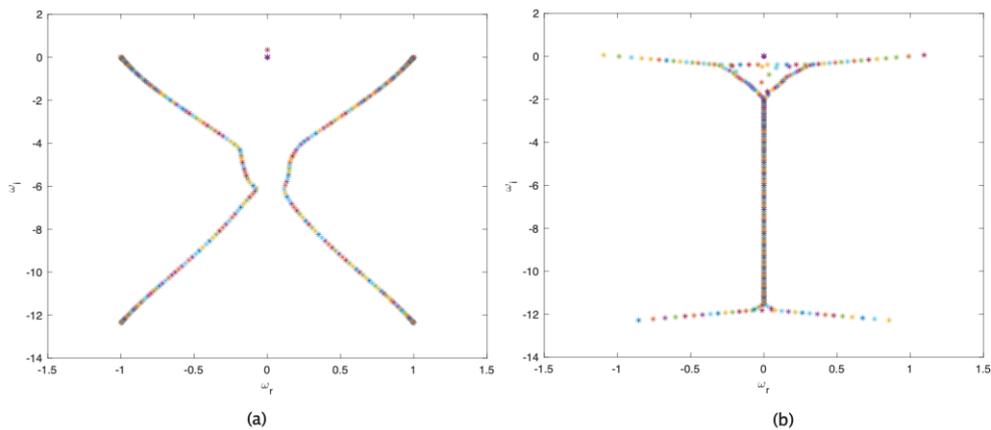

Figure 6. The eigenvalue based on linear stability theory for parallel flow. a) the temporal mixing layer flow; b) the Couette flow.

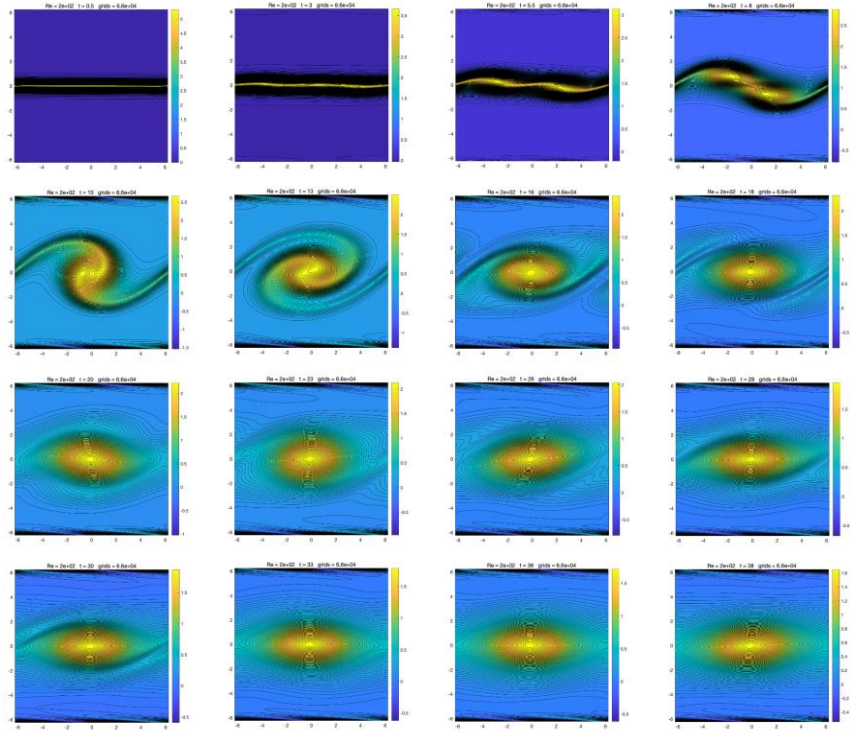

Figure 7. The evolution of the main vortices in the mixing layer from the beginning to the merging under the condition of Reynolds number of 200, in the order of left to right and top to bottom with a time interval of 2.5s.

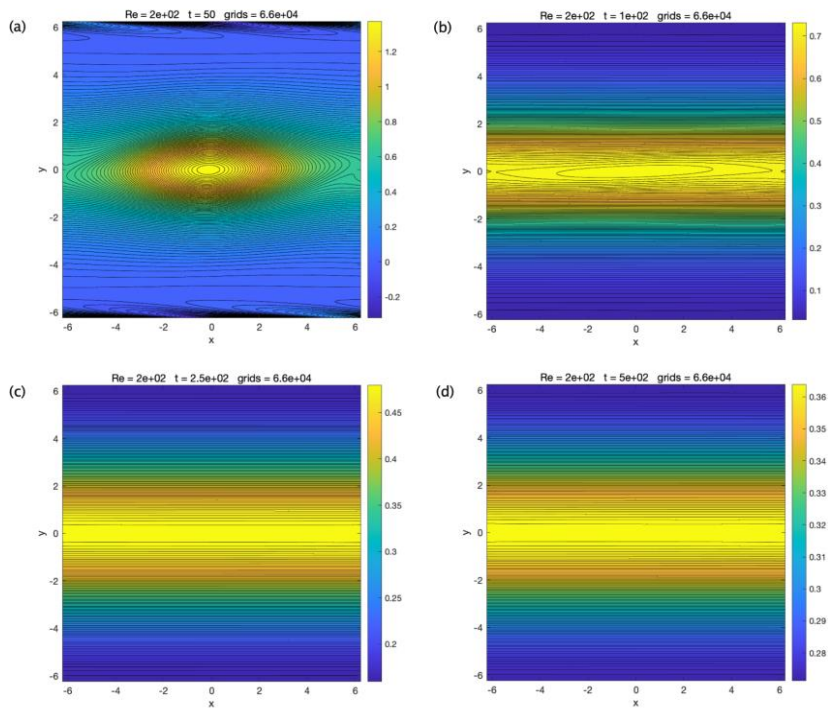

Figure 8. The evolution of the main vortices in the temporal mixing layer from the merging to the dissipation and degeneration under the condition of Reynolds number of 200.

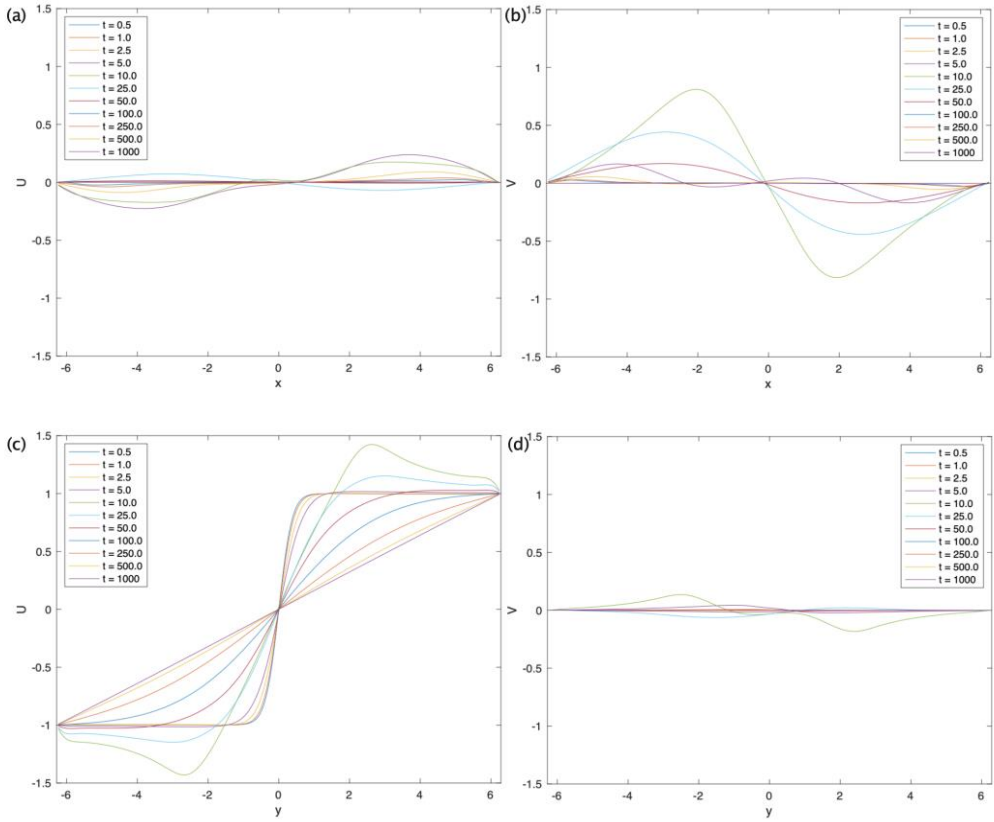

Figure 9. the evolution of velocity in the central horizontal and vertical directions over time under the condition of Reynolds number of 200. a) the relationship U-x; b) the relationship of V-x; c) the relationship of U-y; the relationship of V-y.

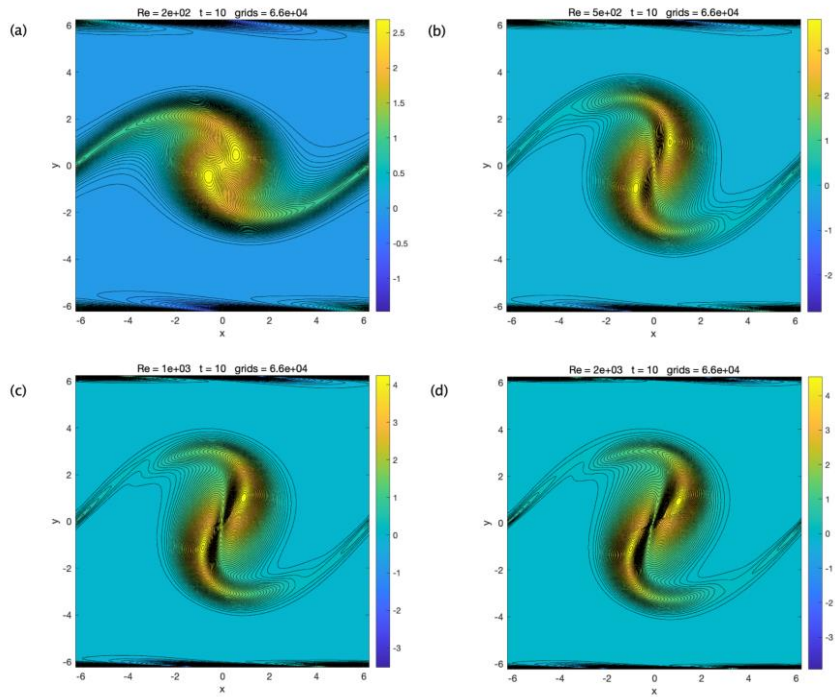

Figure 10. The effect of Reynolds number on the evolution of vortices in mixing layer at intermediate Reynolds numbers (t=10s). a) $Re = 200$; b) $Re = 500$; c) $Re = 1000$; d)

$Re = 2000$.

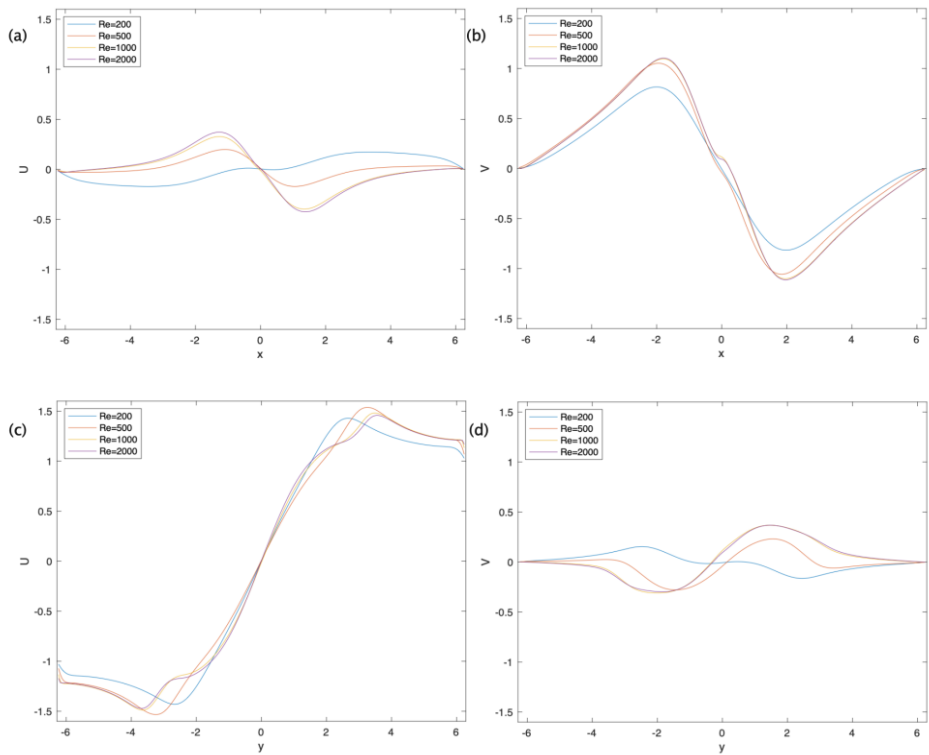

Figure 11. The velocity distribution in the central horizontal and vertical directions under different Reynolds numbers from 200 to 2000 (t=10s). a) the relationship U-x; b) the relationship of V-x; c) the relationship of U-y; the relationship of V-y.

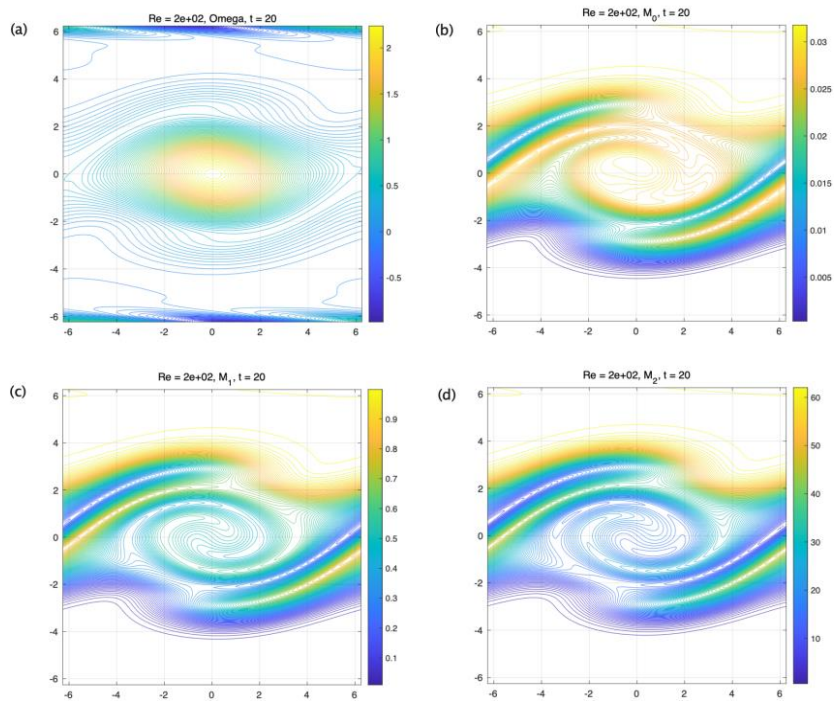

Figure 12. The distribution of particle moments under the conditions $Re = 200, Sc = 1, Da = 1$ with AK-iDNS framework (t=20s). a) Vorticity; b) $M_0$; c) $M_1$; d) $M_2$

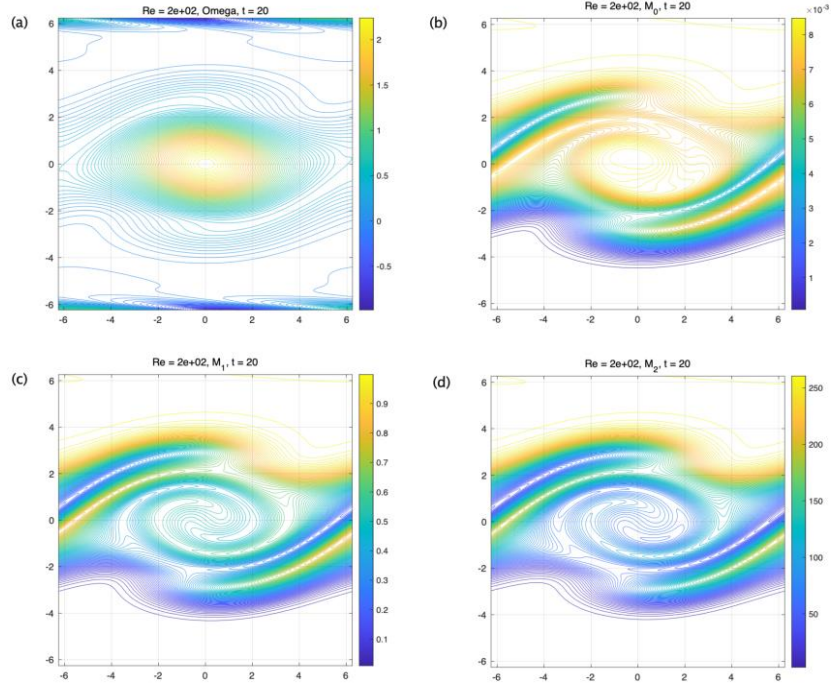

Figure 13. The distribution of particle moments under the condition $Re = 200, Sc = 1, Da = 1$ with TEMOM (t=20s). a) Vorticity; b) $M_0$; c) $M_1$; d) $M_2$

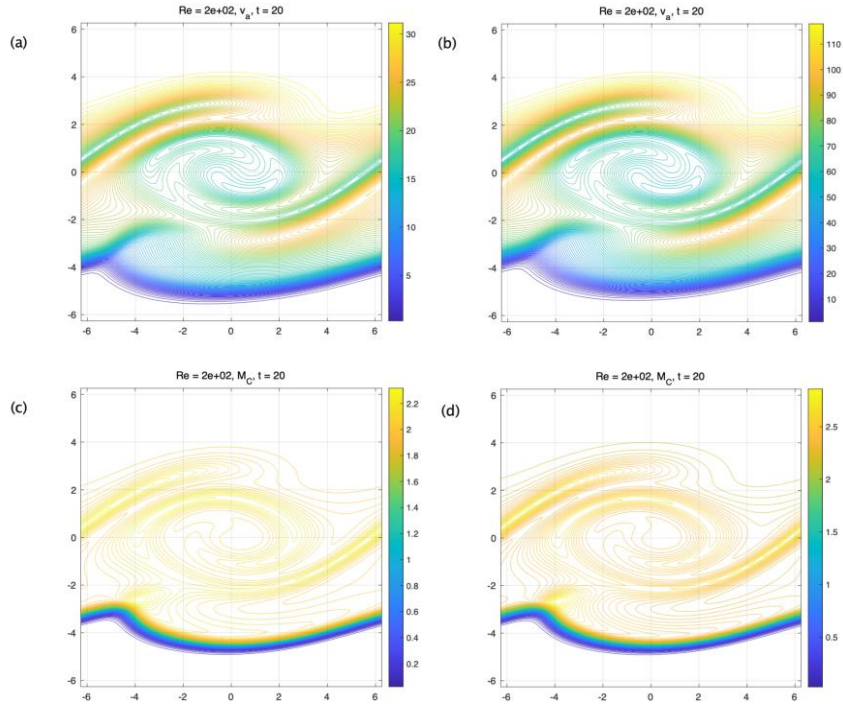

Figure 14. the comparison of the distribution of algebraic mean volume and dimensionless particle moment between AK-iDNS and TEMOM under the conditions $Re = 200, Sc = 1, Da = 1$ (t=20s). a) algebraic mean volume with AK-iDNS; b) algebraic mean volume with TEMOM; c) dimensionless particle moment with AK-iDNS; d) dimensionless particle moment with TEMOM.

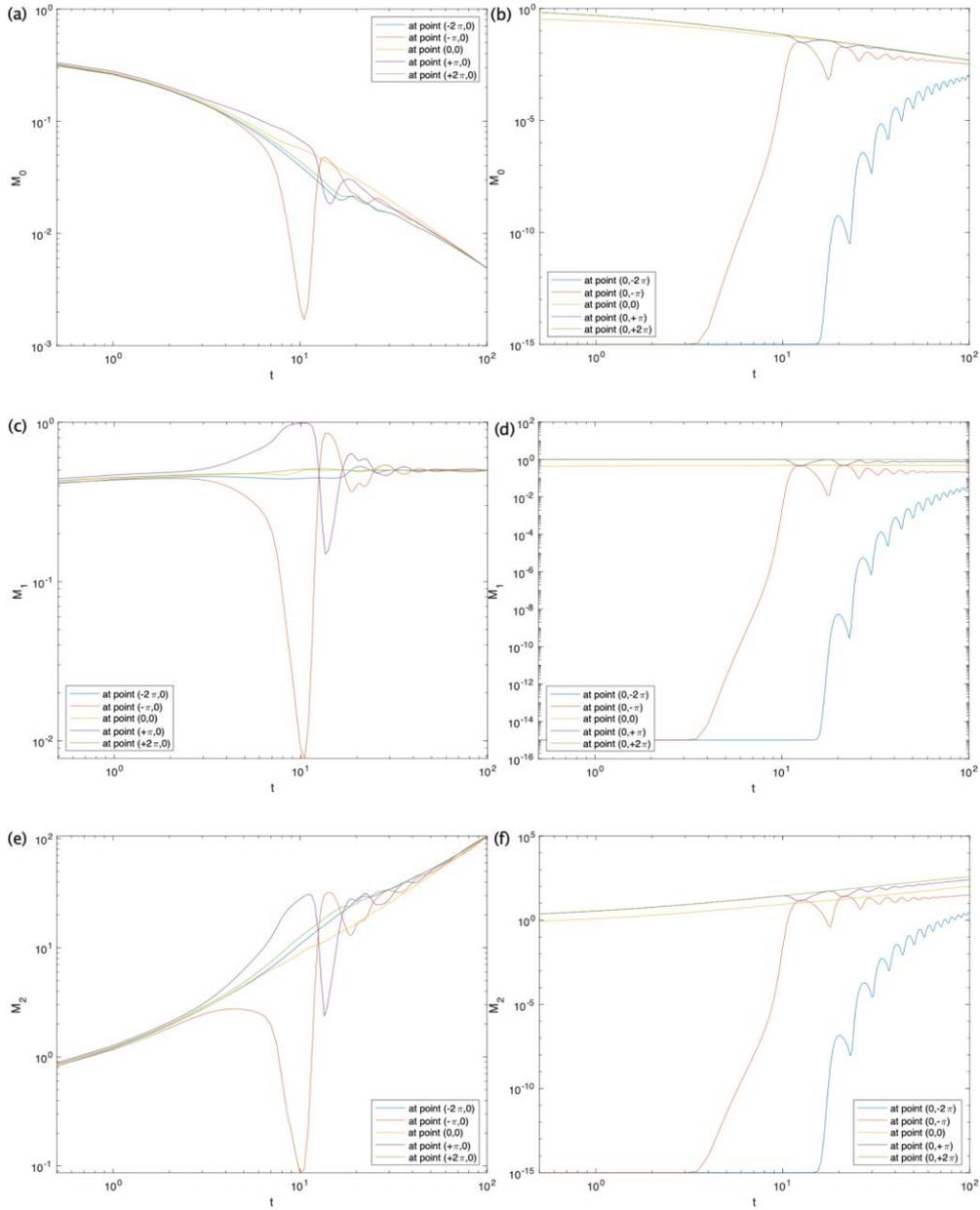

Figure 15. The evolution of particle moments with time at different point under the conditions $Re = 200, Sc = 1, Da = 1$. a) $M_0$ along the central horizontal line; b) $M_0$ along the central vertical line; c) $M_1$ along the central horizontal line; d) $M_1$ along the central vertical line; e) $M_2$ along the central horizontal line; f) $M_2$ along the central vertical line.

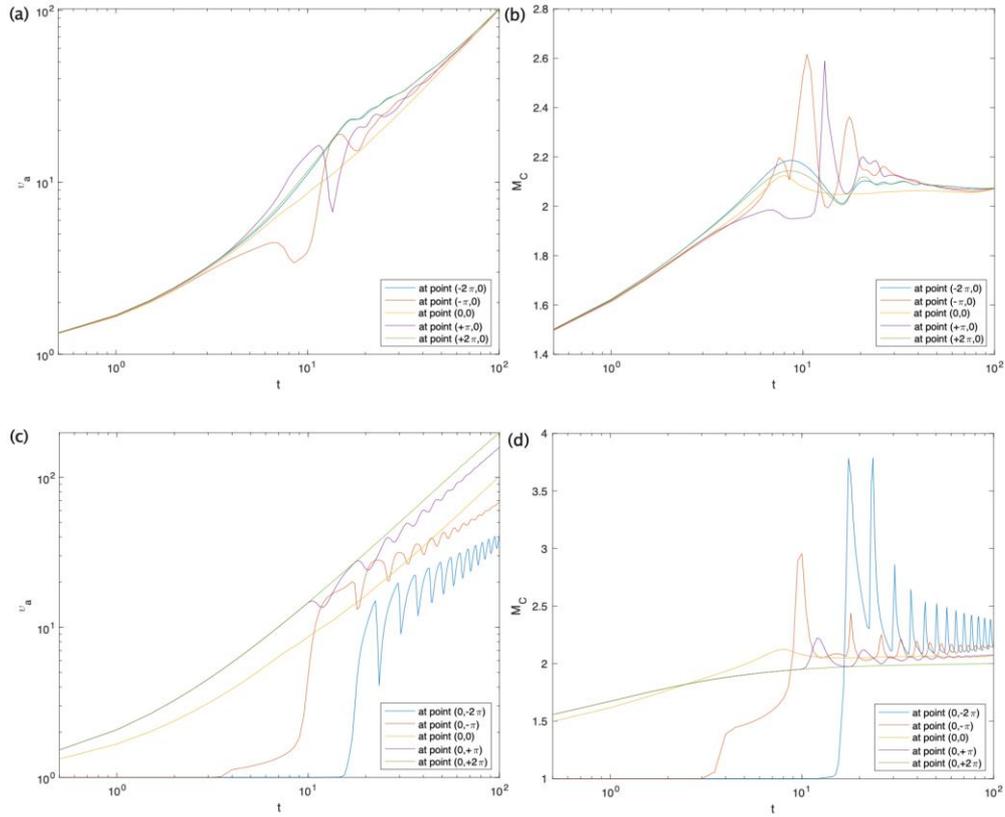

Figure 16. The evolution of particle algebraic mean volume ($v_a$) and dimensionless moments ($M_C$) with time at different points under the conditions $Re = 200, Sc = 1, Da = 1$. a) $v_a$ along the central horizontal line; b) $M_C$ along the central horizontal line; c) $v_a$ along the central vertical line; d) $M_C$ along the central vertical line;

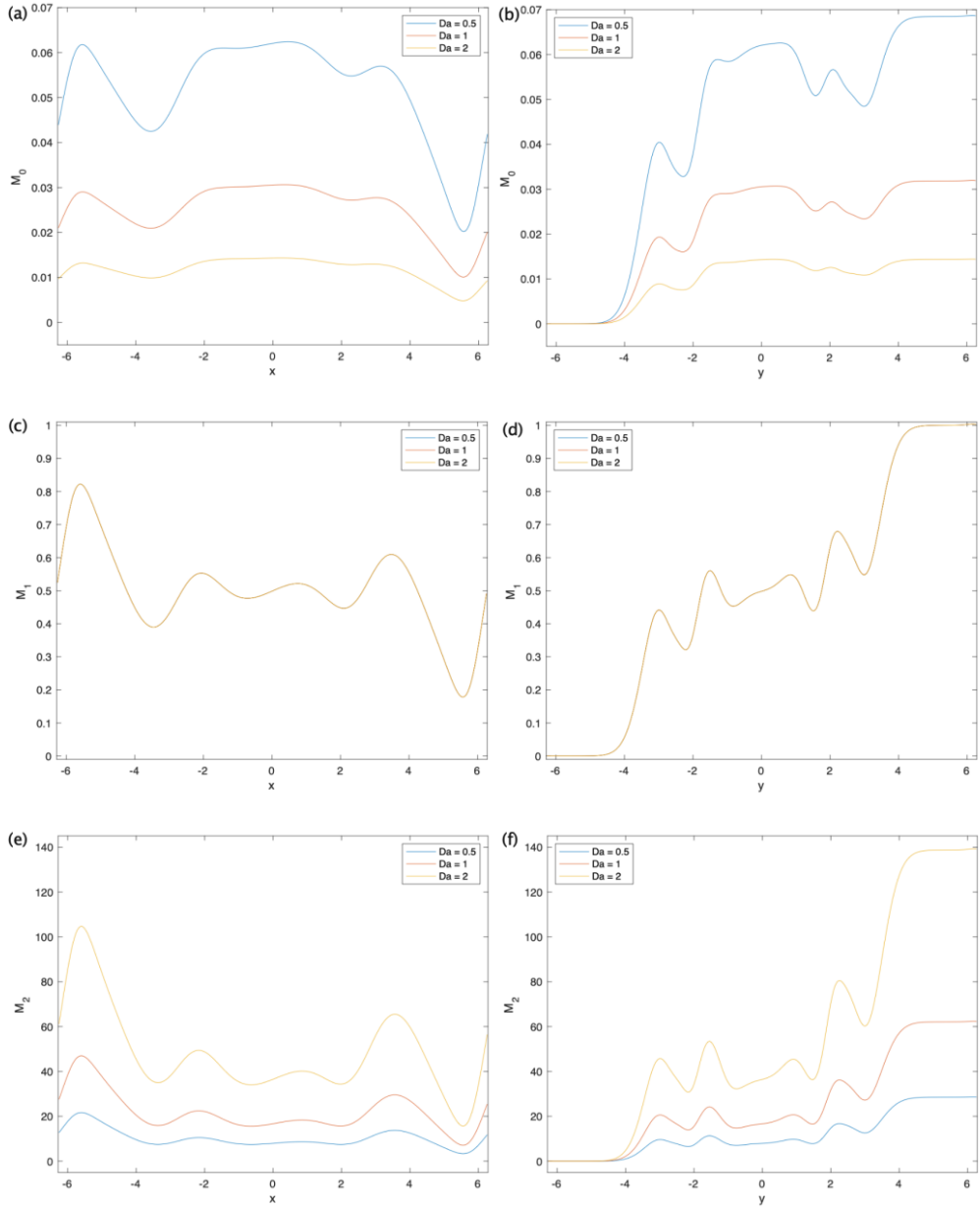

Figure 17. The effect of Damkohler number on the distribution of particle moment in space under the conditions $Re = 200, Sc = 1$. a) $M_0$ along the central horizontal line; b) $M_0$ along the central vertical line; c) $M_1$ along the central horizontal line; d) $M_1$ along the central vertical line; e) $M_2$ along the central horizontal line; f) $M_2$ along the central vertical line.

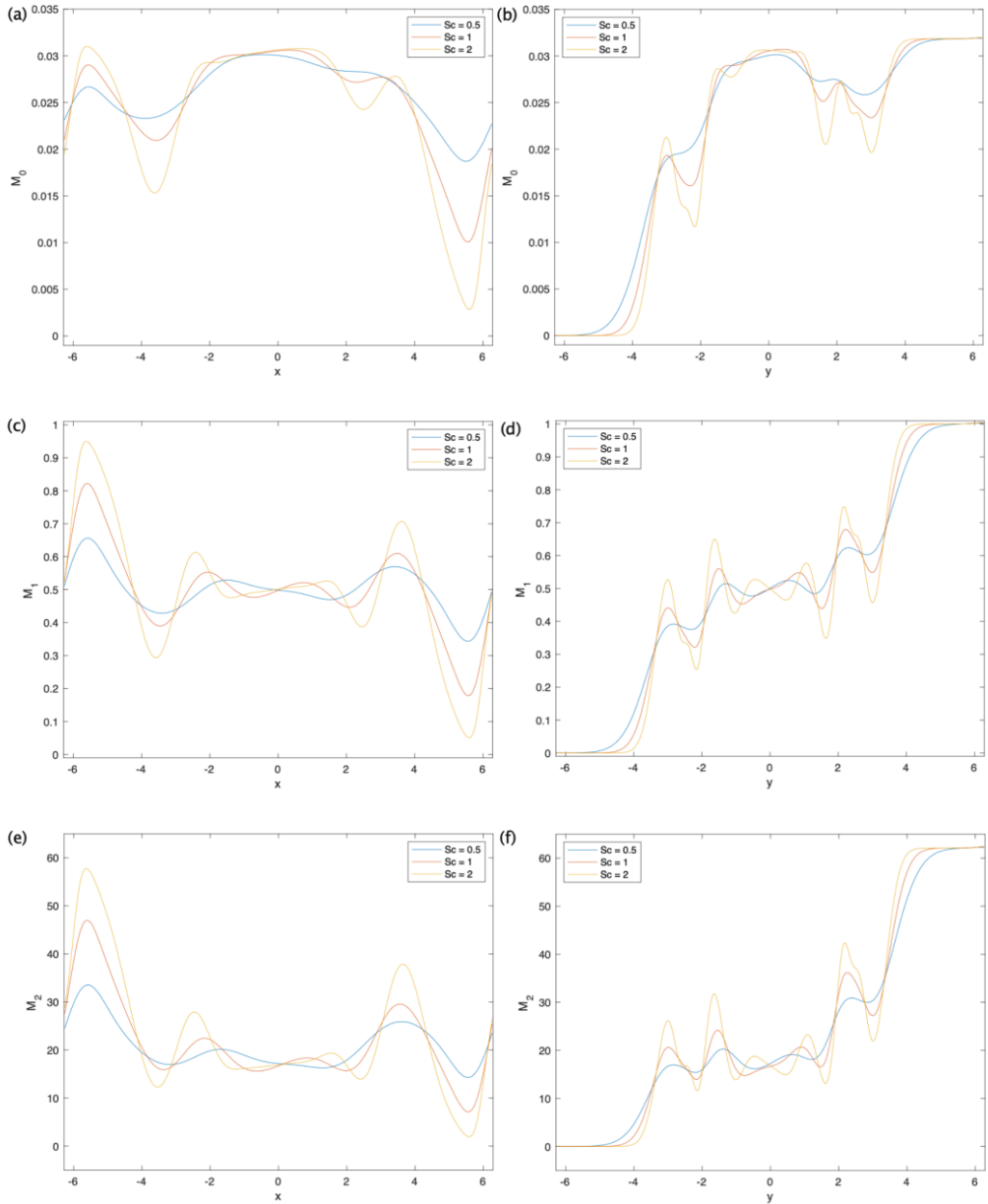

Figure 18. The effect of Schmidt number on the distribution of particle moments in space under the conditions $Re = 200, Da = 1$. a) $M_0$ along the central horizontal line; b) $M_0$ along the central vertical line; c) $M_1$ along the central horizontal line; d) $M_1$ along the central vertical line; e) $M_2$ along the central horizontal line; f) $M_2$ along the central vertical line.

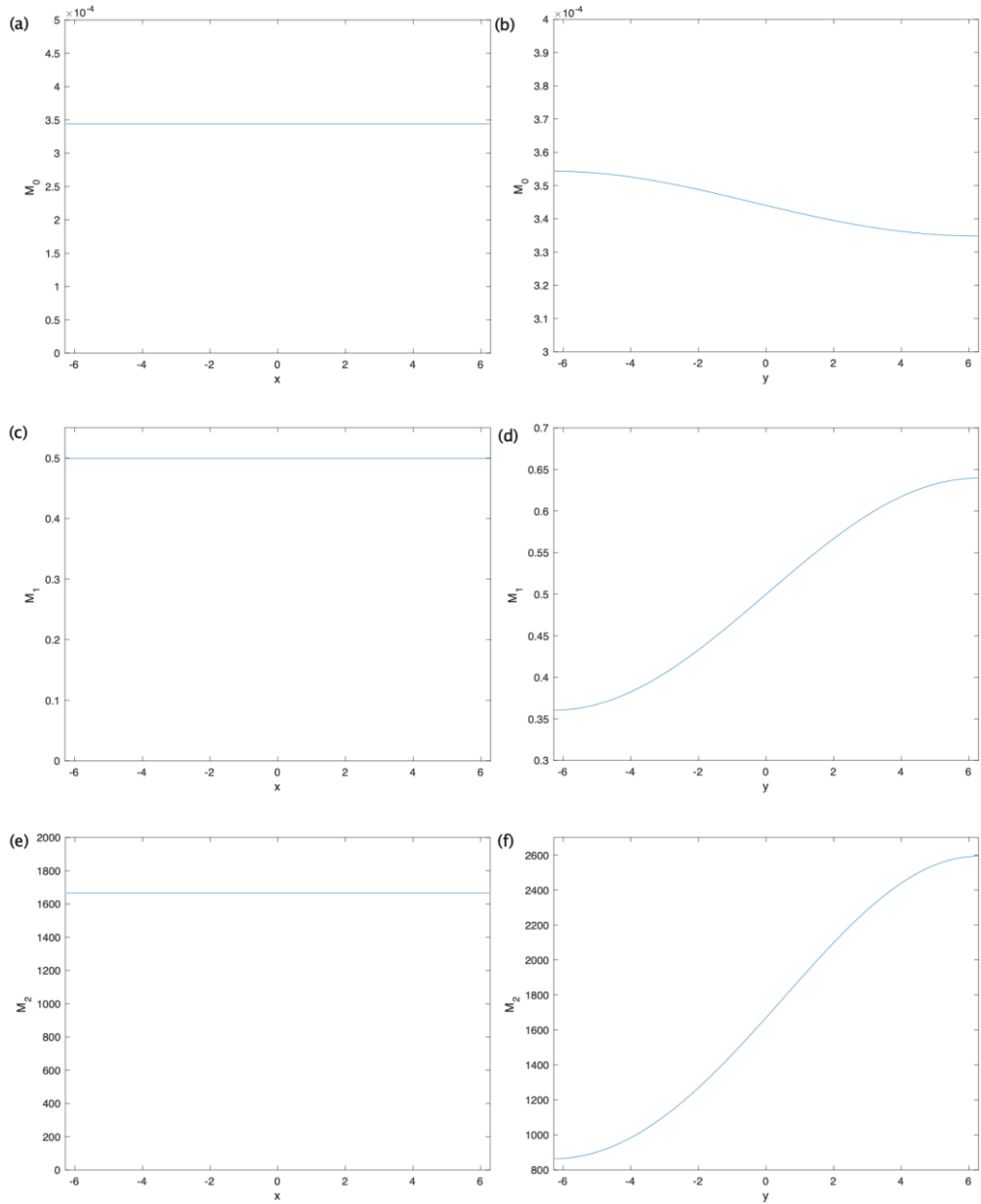

Figure 19. The asymptotic distribution of particle moments in space under the conditions $Re = 200$, $Sc = 1$, $Da = 1$ at $t = 1000$ s. a) $M_0$ along the central horizontal line; b) $M_0$ along the central vertical line; c) $M_1$ along the central horizontal line; d) $M_1$ along the central vertical line; e) $M_2$ along the central horizontal line; f) $M_2$ along the central vertical line.

**Declaration of interest statement**

We declare that we have no finical and personal relationship with other people or organizations that can inappropriately influence our work. There is no professional or other personal interest of any nature or kind in any product, service and company that could be construed as influencing the position presented in, or review of, the manuscript entitled.

Mingliang Xie

State Key Laboratory of Coal Combustion, Huazhong University of Science and Technology, Wuhan 430074, China

Correspondence Email: mlxie@mail.hust.edu.cn